# From a set of parts to an indivisible whole. Part I: Operations in a closed mode

Leonid Andreev

Equicom, Inc., 10273 E Emily Dr, Tucson, AZ 85730, U.S.A.
E-mail: equicom@matrixreasoning.com

February 29, 2008

**Abstract**

This paper provides a description of a new method for information processing based on holistic approach wherein analysis is a direct product of synthesis. The core of the method is iterative averaging of all the elements of a system according to all the parameters describing the elements. It appears that, contrary to common logic, the iterative averaging of a system's elements does not result in homogenization of the system; instead, it causes an obligatory subdivision of the system into two alternative subgroups, leaving no outliers. Within each of the formed subgroups, similarity coefficients between the elements reach the value of 1, whereas similarity coefficients between the elements of different subgroups equal a certain constant value of $0>\Omega<1$. When subjected to iterative averaging, any system consisting of three or more elements of which at least two elements are not completely identical undergo such a process of bifurcation that occurs non-linearly. Successive iterative averaging of each of the forming subgroups eventually provides a hierarchical system that reflects relationships between the elements of an input system under analysis. We propose and discuss a definition of a natural hierarchy that can exist only in conditions of closeness of a system and can be discovered upon providing such an effect onto a system which allows its elements interact with each other based on the principle of self-organization. We show that self-organization can be achieved through an overall and total cross-averaging of a system's elements. We propose an algorithm for performing such cross-averaging through iterative averaging transformations of a system's similarity matrix, wherein the very first of the iterative transformations turns any system under processing into a closed type system that does not allow an addition of new elements or removal of any of its existing systems as it would result in drastic changes as compared to the original state of the input data system. A system subdivision into groups occurring in the course of iterative averaging performed in an autonomous unsupervised mode displays a highly intelligent analysis of part-whole relations within the system, which proves that the resulting hierarchical structures reflect the system's natural hierarchy. This method for data processing, named by us 'matrix reasoning', can be effectively utilized for analysis of any kind and any combination of data. We demonstrate new methods for construction of hierarchical trees, dendrograms, and iso-hierarchical structures which allow effective visualization of results of a hierarchical analysis in the form of a holistic picture. We demonstrate the application potentials of the proposed technology on a number of examples, including a system of scattered points, randomized datasets, as well as meteorological and demographical datasets.

## 1. Introduction

Part-whole relations are one of the structuring bases of the universe. One may assume that since the ancient times the problem of 'part-whole' relations has been most stimulating for development of philosophical understanding of the nature of human-being and the environment. The principle of approaching a whole from the standpoint of its parts and treating the properties of a whole as the sum of properties of its parts is known as *merism* (from the Greek *'meros'*, 'part') and is the subject of studies in mereology [1-3]. Alternatively, a position which emphasizes the inequality between a whole and the sum its parts because a whole, due to its parts, acquires new properties as compared to its parts, is known as holism (from the Greek word *'holos'*, 'whole') [4-5]. Holism is based on the idea that all properties of a given system cannot be determined or explained by the sum of its component parts alone. Instead, the system as a whole determines in an important way how the parts behave. Holism as a philosophical paradigm is aimed at a "holistic perception of the world", i.e. at resolving the conflict between the subjective and objective, between irrational and rational. The holistic approach concerns all the areas of philosophy as it deals with the general principles of scientific discovery of knowledge. 'Wholeness' as a display of the properties of a subject under investigation, hence the entire cognizable world, is viewed by holistic science not as something that directly and obviously follows from interrelations between the elements of a system, but as something that is manifested in the existence of specific and stable properties of such wholeness. Holism maintains that one of the fundamentally important properties of a whole is nonreducibility of the properties of a whole to the properties of its component parts, which is in contrast to analytic tradition of establishing the properties of a whole through analysis of cause-and-effect relations and relationships between the parts of a whole which have a fixed set of properties. From the standpoint of holism, the requirement of logical deducibility of properties of a whole from initially set conditions, as it is maintained by traditional scientific methodologies, is the reason of why all the diverse and extensive domains of knowledge about Nature's objects and phenomena still fail to provide a holistic picture of the world. Extreme forms of the concept of holism exist largely due to unavailability of scientifically grounded methods – or even ideas that would promise a potential capability of development of such methods – for synthesis of a whole which could provide that a resulting whole, rather than being a sum of the component elements, would acquire new properties that were not present in the component elements.

Analysis of the behavior of parts from the viewpoint of the whole is not characteristic of the classical science. In classical science, analysis is based on breaking down a phenomenological whole into its parts and examining the parts, and this reductionist process is not complemented with a reverse process – from parts to a whole – and therefore it does not provide an integral picture of a system under analysis. The problem of relationships between real, identifiable subjects and the appearances whose emergence supposedly involves participation of those subjects seems to be the most critical and complicated issue of the modern scientific knowledge. As was emphasized by Craig Dilworth [6], "The debate over empiricism and realism concerns the very nature of modern science: what it is or what it ought to be. Empiricism, in its extreme form, claims that there is no reality behind appearances and that it is the task of science to determine what the appearances are and what the formal relation are that obtain among them." A functional whole may contain any kind of elements (elements are parts of a system which do not consist of subsystems, and the notion of 'element' includes also peculiarities of an element's interactions with other elements of the whole), even something that is unknown to science and lives and evolves on its own, independently from us, the humankind. Notwithstanding the overwhelming amount of publications on philosophical understanding of the problem of 'whole-part' from the standpoint of holism, science does not know of mathematical ideas and exact methods that would be able to demonstrate, on the quantitative level, the relations between parts which underlie the functioning of an integral indivisible whole. This problem is profoundly important and hardly solvable in general from the position of linear logic prevailing in science. Kurt Koffka, one the classics of Gestalt psychology, wrote, "It has been said: The whole is more than the sum of its parts. It is more correct to say that the whole is something else than the sum of its parts, because summing up is a meaningless procedure, whereas the whole-part relationship is meaningful." [7]. Eighty years ago, Jan Smuts, who coined the term 'holism' and made an important contribution to the philosophy of holism, outlined the problem that would be faced by the science were it to attempt a description of holism: "This is … the case where cell *a* unites with cell *b* to form a new entity, in which both *a* and *b* disappear finally and irrecoverably, and whose character and behaviour cannot be traced mathematically or mechanically to those of *a* and *b*." [8]. It is "…impossible to say where the whole ends and the parts begin, so intimate is their interaction and so profound their mutual influence" [9].

Due to its universality, the problem of whole-part relations is inexhaustible for analysis and understanding. A. J. Bahm [10], for instance, points out five kinds of whole-part relations: atomism, holism, emergentism, structuralism and organicism, and describes the relations between them. The diversity of relations between a whole and its parts is discussed in [1] as well as in other sources. Scientific systematics of whole-part relations would require a capability to identify at least statistically verifiable gaps between subsystems or the existence of some other criteria for differentiation between subsystems. However painful it might be for the scientific community to admit it, the systematics of whole-part relations becomes hardly feasible as one moves forward from additivity towards wholeness. The objective reason for this problem lies in the fact that the 'whole-parts' theory, especially in the part concerning an indivisible whole, is essentially based on notions – such as systemity, metaphysics, evolution, hierarchy, chaos, cooperative relations, etc. – that are all-embracing and difficult for interpretation, not only quantitatively, but qualitatively, too.

Advances in the theory of 'parts-whole' relations will greatly impact the science and technologies of the future, particularly, such complex and controversial areas of science as artificial and natural intelligence, risk assessment, modeling of unpredictable situations, catastrophe theory, statistics and economics, medical and social psychology, including collective behavior, quantum physics, astrophysics, and many others. Solutions for many critical problems of the nowadays science directly depend on novel approaches to holistic processing of information. There has to be a certain universal methodology providing a capability to objectively evaluate a given set of elements from the point of view of its ability to become – upon certain variations in conditions of their interactions – a non-additive, indivisible, specific entity, i.e. a phenomenon whose scientific analysis allowed for distinction of those elements. Currently, a determination on how relevant is a given set of subjects in the emergence of an indivisible whole under analysis is made by using specific and subjective approaches that are based on and result from a combination of such factors as experience, opinions and beliefs of the analyst or a team of analysts, as well as methods in mathematical statistics which are completely incompatible with the paradigm of holism.

The purpose of this research was the development of a universal methodology for synthesis – from analytically discovered and independently identified and described elements of a functional whole – of the initial intact whole that possesses stable and specific properties that are not present in either its individual component elements or a mechanical totality of those elements. The said methodology utilizes a new original data processing technology based on iterative averaging, over all available parameters, of each and all elements of a system under analysis [11]. As a result of the iterative averaging, all the elements of the system divide into two subgroups, without any outliers or transient elements. The absence of transient elements is an important peculiarity of this technology as it completely excludes the possibility of subjective interpretation of results and allows for unsupervised autonomous data processing. A dichotomy resulting from the iterative averaging provides a 100% similarity of elements within a subgroup, whereas a similarity between the subgroups may widely vary depending on initial properties of the system's elements, and it does not affect the division into subgroups. The iterative averaging provides an evolutionary transformation of the system under analysis, involving both convergence and divergence processes. Successive evolutionary transformations of each of the emerging subgroup of elements eventually produce a hierarchical tree that shows the properties of the whole, which are not present in any of its individual elements or in a mechanical totality of those elements. Thus, while reductionist methodology offers a one-way path to knowledge by following from the root of a hierarchical tree to its leaves, and while any network is a certain form of presentation of a hidden hierarchy, namely a "leaf distance matrix" whose informational potential dramatically decreases in conditions of dynamical existence of a given hierarchical system, the proposed methodology of construction of hierarchical trees by analyzing the properties of their leaves can be viewed as a new paradigm of epistemology, congruent with the paradigm of holism.

The technology described in this paper involves three principally different approaches to synthesis of a whole from a totality of its component parts, and therefore the paper consists of three parts: Part I describes "closed mode operations", Part II deals with "open comparative mode operations", and Part III provides a method for analysis of holistic space of multi-object relations. In case of a closed mode, the very first step of information processing with the use of evolutionary transformation of similarity matrices turns the totality of the initial elements (parts) of the system into a closed system wherein the number and quality of elements should not and cannot change. In case of the open comparative mode, each individual element of the system is consecutively compared to a certain outside element (i.e. not belonging to a given system) or a set of outside elements that carries a certain meaning for the analyst. The technique for such comparison, comprising the algorithm of evolutionary transformation of similarity matrices, represents in general a system of algorithms named by us as "information thyristor" [12]. Essentially, this technique enables the analyst to establish on a quantitative level whether or not a certain idea may apply to a given set of elements of a system, and thus it provides a new approach to hypothesis generation and verification. Part III describes a methodology that involves the use of an outside "drifter" object that is chaotically and unrestrictedly moves within the system's space while being compared, at its every move, to all of the elements of the system. Coupled with the method of evolutionary transformation of similarity matrices, the use of a drifter provides unique information about interactions between the elements of a system and, in particular, allows the assessment of the size and shape (i.e. aura) of the system's space beyond which the intra-system interactions cease to exist. It appears that intra-system interactions are described by a very complex structure of closed attractor membranes which depends on parametric characteristics of the system's elements and the number of those elements.

It would be impossible within the framework of a single article to thoroughly discuss those aspects of evolution, metaphysics, hierarchy and systemity which are directly connected with the problem of 'part-whole' relations, hence with the technology presented in these papers. Nonetheless, in order to better understand the proposed technology and its implications for fundamental and applied science, it will be necessary to touch upon some of the above-said issues. In this paper, we will consider some of the theoretical problems that are important in the context of the technologies presented in this series of three articles.

## 2. Holistic perception of information and construction natural hierarchies

### 2.1. Evolution

Computer science deliberately uses the "evolutionary" epithet, such as in 'evolutionary computation', 'evolutionary algorithms', 'evolutionary programming', etc. This comes primarily from a desire to state that a certain software product is not merely an additive set of commands based on linear logic but a system capable of independent development and self-organization. In this context, we would like to point out a certain criterion that can indicate whether or not a given computer program indeed can perform actions similar to Darwinian natural selection.

In Chapter 6 of "The Origin of Species", titled "Difficulties on Theory", Charles Darwin remarked that there were some problems that, in his opinion, could be fatal to his theory: "These difficulties and objections may be classed under the following heads: Firstly, why, if species have descended from other species by insensibly fine gradations, do we not everywhere see innumerable transitional forms? Why is not all nature in confusion instead of the species being, as we see them, well defined?" [13]. In Darwin's time, the science of paleontology was yet at the embryonic stage of its development and the fossil record was poorly known, so Darwin could only hope that in the future the situation would change for the better. However, within the past 150 years, not much has changed in that respect. "One of the most surprising negative results of paleontological research in the last century is that such transitional forms seem to be inordinately scarce. In Darwin's time this could perhaps be ascribed with some justification to the incompleteness of the paleontological record and to lack of knowledge, but with the enormous number of fossil species which have been discovered since then, other causes must be found for the almost complete absence of transitional forms." [14]. Thus, what has been proven on a large number of examples should be accepted as an axiom or law: the most important point of Darwinian natural evolution is that species formation represents a discontinuous function or – more precisely – a process of sharp dichotomization.

It clearly follows from the above that the adjective "evolutionary" in reference to a computer program must be used responsibly. To claim that a certain computer-based procedure for data processing has semblance to a natural process of evolution, the "novel and coherent structures, patterns and properties" [15] arising in the course of data processing must be free from transitional forms. The fact that the iterative averaging of properties of a system's elements, as described below, leads to the system's sharp dichotomy with no transient elements, makes it fully conform to the criterion of natural evolution. Reconstruction of phylogenesis, i.e. moving from the leaves of a hierarchical tree towards its root, as provided by the algorithm of evolutionary transformation, consists in successive averaging of the properties of antipodes that evolved from their predecessor and represents a perfect illustration of a holistic perception of evolutionary processes.

**2.2. Metaphysics**

"The question of holism must be approached from a metaphysical point of view: as the task of determining the level up to which a property is constituted by its relation to other properties" [16]. Indeed, metaphysics as a branch of philosophy which studies the nature of the universe as a whole science of being and knowing is fully compatible with the paradigm of holism since any doctrine that emphasizes the priority of a whole over its parts is holism. Metaphysics that deals with fundamental problems of Weltanschauung has always influenced concrete science as no single scientific theory can be tested in isolation. It seems to be quite natural that metaphysics must have emerged and developed due to an expressed inability of the average human mind to perceive new knowledge in a way other than viewing it as an additive construction based on the existing knowledge. At the level of linear logic, it is impossible to perceive the universality of regularities of formation of antipodes within a system whose elements undergo iterative averaging, and this seems to be the most clear and compelling proof of the legitimacy and validity of the metaphysical view of the surrounding world. Therefore, analysis of the history of relationships between metaphysics and specific fields of science can help see the tendencies in the development of methodology of scientific knowledge with regard to the 'parts-whole' problem.

Unlike science, metaphysics strives to arrive to an ultimate and overall perspective from which it would be possible to explain all the aspects of existence as it is. Therefore, metaphysical knowledge is more conservative and hard to refute, hence more stable than scientific knowledge. As the fundamentals of scientific knowledge get more complex, the finding of formal relations between appearances becomes more complicated due to limitations in the human mind capabilities and potentials, whereas the need in visualization capabilities upon demonstration of scientific results that go above and beyond the frames of strictly scientific relations is constantly growing. Science cannot exist without criteria of veracity of knowledge. During smooth periods of the development of science, the criteria of veracity are based on the scientific Weltanschauung supported by academic schools of thought and scientific authorities. However, in the times of its turbulent development, science leans on metaphysics for criteria of veracity of the new knowledge. The late 19$^{th}$ and especially the early 20$^{th}$ centuries were the time of unparalleled revolutionary discoveries in almost all fields of science, as well as of unprecedentedly high rate of development of theoretical knowledge and precision engineering. As a consequence, during the recent decades, science has shown a global tendency toward integration, or rather eclectic combination, of scientific and philosophical knowledge, or – to be more precise – toward imitation of metaphysical approaches and methods, which has produced such new areas of science as the theory of self-organization [17], the general theory of systems [18], non-equilibrium thermodynamics and the theory of dissipative structures [19], synergetics [20], fractal geometry [21], string theory [22], as well as some new notions, such as complexity [23], emergence [15], etc. The emergence of these quasi-metaphysical theories and terms is both natural and symbolical, which has been already commented on in philosophical publications. J. W. N. Watkins [24] noted, "The counter-revolution against the logical empiricist philosophy of science seems to have triumphed: I have the impression that it is now almost as widely agreed that metaphysical ideas are important in science as it is that mathematics is". It should be also emphasized that, for understandable reasons, putting a purely scientific knowledge into a quasi-metaphysical form always stirs enthusiasm in the scientific community but does not lead to perfection of scientific tools and means.

There is something in common in the aforementioned quasi-metaphysical theories, as well as in all other quasi-metaphysical theories. One of their common features is that they represent a non-controversial and logical expansion of science to a scale that is so large that it seems to be a metaphysical scale, even if it has nothing to do with metaphysics which studies what lies beyond the boundaries of physical phenomena. Thus made artificial stretching of scientific knowledge ought to be non-controversial in order to avoid outright rejection. In other words, it cannot afford being radically novel or too contradictory to commonly accepted scientific paradigms. For example, after Haken's monograph [20], the concept of synergetics can be effectively used in scientific discussions in any field of science. The term 'synergetics' is known to fit any scientific topic, without an immediate risk of its improper use. It is a convenient word to use when speaking of the universe as a complexly organized live entity in constant dynamical and evolutionary self-development, but, in fact, all it represents is just a way of suprascientific interpretation of reality and essentially is profanation of metaphysical knowledge. The synergetics-based way of thinking is supported mainly by generalized symbols of certain, yet to be acquired knowledge, nonlinear logic and other attractants of idealized science, rather than by concrete particulars of the nonlinear world which could be studied today and to the point.

Another quasi-metaphysical theory – the superstring or M-theory [22] – has become a part of elementary particles physics. In certain conditions, superstrings may reflect the properties of ordinary particles, but at the discretion of a physics theoretician – for instance, upon the increase of dimensionality – they may become extremely complex and acquire most unexpected and unpredictable properties and thus stimulate the researcher to look at an old problem in a new way since a traditional approach to the problem turns out to be too complicated. According to P. Woit, a quantum field theorist, currently working in mathematics [25], no one has ever been able to make any experimental predictions based on the string theory, and there is a legitimate question of whether the string theory is a scientific theory at all. In Woit's opinion, the only area in which the string theory is really strong is public relations (which is true for all and any other quasi-metaphysical theories, and which distinguishes the latter from the true metaphysical outlook that has always had tremendous problems in the area of public relations). Woit, admitting that the leading theorists in the string theory were undoubtedly geniuses, has thoroughly analyzed the reasons of why the string theory had caused such overwhelming enthusiasm in quantum physics. It was quite understandable as the disproportionately drastic formal expansion of science brought about high expectations in respect of many problems that theretofore seemed to be irresolvable. First and foremost, it applies to the problem of quantum gravitation theory that has always attracted the most talented physicists. Another example of quasi-metaphysical approach based on an unlimited non-controversial expansion of theoretical knowledge is fractal geometry effectively used by B. Mandelbrot for development of reproducible techniques in recognition of indefinitely expanding elements of fractality in various mathematical and natural forms [21].

The use of the technique of non-controversial formal expansion of theoretical knowledge could be also demonstrated on examples from all of the above-named areas of science which exploit the imitation of the metaphysical level of knowledge. This has become a stable tendency in modern science, and it has a strong motivational effect on scientists but very limited potentials in producing new concrete knowledge about objects under study, unless, of course, 'new knowledge' means new ways of presentation of widely known truths. It is easy to prove that the more elaborate are the attempts to tightly bound scientific knowledge with metaphysics, the slower is the progress in fundamental scientific knowledge, and the faster is the development of applied sciences where fundamental knowledge, despite its overall positive role, has an inhibiting effect on innovativeness.

Notwithstanding the congruence between metaphysics and holism, they pursue fundamentally different goals in acquisition of knowledge. Unlike metaphysics, the paradigm of holism is aimed at concrete knowledge, and therefore there is no such thing as quasi-holistic science. If a union between holism and specific fields of science will ever be built, it may exist only based on exact knowledge, and therefore the prefix "quasi" will not be applicable. The emergence of quasi-metaphysical theories clearly demonstrates that the ever-existing balance between holism and greedy reductionism has a strong tendency of shifting towards holism.

### 2.3. The concept of system

The term 'system' is one of the most widely used terms both in science and life, not only due to its semantic plasticity, but also because it reflects some of the very important universalities of being. Among the numerous definitions of 'system', the key words are a "united whole" wherein all of the constituent parts work together to perform a certain concrete function. Hence, the inevitably following conclusion is: any functioning system determines "by itself" what it needs in order to perform its function, and what it does not need. If a certain element of the system does not need to be engaged at a certain given moment of time, but it will be needed at a later time or upon variations in the system's environment, that element should necessarily be considered as a systemic element, as without it the system would lose its functional potentials. Thus, all things considered, a functioning system is a closed formation, even if it includes, along with obligate elements, facultative ones, which creates certain diffusion of a conventional border between the system and outer environment, and therefore a word combination "open system" is nonsense. From the standpoint of holism, elements that are responsible for a system's functioning as a whole are the system's immanent parts. "Open" or "closed" should rather refer to the operational mode applied to analysis of a system, but not a system itself which by definition can only be a closed formation.

The "independence" of functioning systems was conceptually reflected in one of the earliest quasi-metaphysical theories – W. R. Ashby's theory of self-organization [17] in regard to processes in which dynamic functional systems becomes more organized over time and on their own, i.e. without management by outside agents. The concept of self-organization – or "order for free", as was well put by S. Kaufman [26] – is one of the most difficult issues in modern scientific knowledge as the theory of self-organization is essentially a collection of observations, precedents and assumptions, rather than a scientific hypothesis providing instructions on how to discover concrete mechanisms of self-organization in real-world systems. Probably the most impressive results in study of self-organization were obtained by a Nobel laureate Jean-Marie Lehn in his numerous works on supramolecular structure self-assembly (see, e.g. [27]). J.-M. Lehn has convincingly demonstrated that information stored in the covalent framework of the components of a complex mixture of molecules is a source of corollaries, instructions and programs recognized at the supramolecular level. "Self-organization … may be directed by the design of both these components and their mode of assembly, i.e. by the molecular information stored in the components and by the supramolecular processing of this information through the interactional algorithm (the interaction pattern involved)" [28]. One of the factors of J.-M. Lehn's success in study of the processes of self-organization was that he investigated systems with comparatively limited and well studied diversity of interaction between elements (molecules). However, in real-world systems, interactions between elements are overwhelmingly diverse and, as a rule, imperceptible by reductionist methods of investigation. Therefore, most of quasi-metaphysical theories refer to such interactions by using a term "cooperative relations" which is too vague to be a scientific term [17-20].

"Independence" of functioning systems is quite a nuisance for science as it imposes insurmountable limitations. If a system "decides on its own" how to structure itself in order to be able to perform its functions, then a reduction of the system, with a purpose of understanding how it does all that, will make it incapable of "making decisions", hence revealing its secrets. This explains why one of the approaches used by quasi-metaphysical theories, e.g. general systems theory [18], the theory of dissipative structures [19], and synergetics [20], is based on distinguishing between open and closed systems. In physics, an "open system" refers to system that exchanges its substance and energy with the outer environment, whereas a "closed system" means a system that does not allow for such an exchange. Clearly, there is deliberate confusion between two absolutely different problems: (1) how a system acquires substances and energy needed for its functioning, and (2) how it functions, i.e. what mechanisms start the spontaneous interactions between individual elements and subsystems. In the course of study of any functioning systems, it will immediately become clear that these two problems are interconnected in a very complex way and cannot be considered apart from each other. This circumstance is the root of the fundamental difficulty of theoretical understanding of systems behavior. Differentiation between open and closed type systems is an artificial separation of the two aforesaid problems which cannot provide any input to understanding of concrete mechanisms of systems functioning.

Ludwig von Bertalanffy [18] believed that the fact that many characteristics of living systems which are paradoxical in view of the laws of physics was exactly due to their being open systems, as living systems cannot, for instance, live without consumption of oxygen and other substrates from the environment and without releasing the products of their metabolism into the environment. I dare think that to a certain extent this was a psychologically motivated position meant to state that we would not be able to understand the origin of life and the ultimate nature of life processes mostly because any biological system is a functional part of a very complex surrounding world. If we place an obligate aerobic organism in a low oxygen medium, then, irrespective of the way it consumes oxygen – through gills, lungs, or diffusion – the organism's reaction to hypoxia will involve certain evolutionarily determined metabolic shifts that are inherent in a given biological system and are not specific for a particular medium. In conditions of anoxia, an organism, as an independently functioning system, will struggle with a lack or absence of oxygen till the very end by mobilizing its own internal mechanisms, and no changes in the environment can revive the organism post mortem. The environment that provides oxygen for an organism does not participate in the organism's struggle for survival in the conditions of hypoxia. After all, we cannot refer to an airplane as an open system even though it uses an external source of energy and releases the products of fuel combustion into the environment. In Part III of this series of articles

"From a set of parts to an indivisible whole", it will be visually demonstrated that a functioning system cannot be an open system.

According to I. Prigogine [19], systems that display dynamical self-organization are open nonlinear systems that are far from thermodynamic equilibrium and resistant to minor disturbances. An increase of order in such systems is coupled with a decrease of order in their environment. In other words, in a system undergoing self-organization, the entropy continuously increases but it concurrently dissipates or is exported to the environment, thus making the second law of thermodynamics valid for any kinds of situations, especially in case of such dissipative structures as living organisms. As noted by F. Heylighen [29], "the export of entropy does not explain how or why self-organization takes place"; however, that question, crucial for theoretical science, does not seem to have been of major concern to I. Progogine as his works were aimed at pushing the classical thermodynamics to the level of metaphysics and turning it into a universal doctrine of the metaphysical scale so it could explain everything, including the origin of life, the space-time relation, the nature of chaos, and solve many other problems of physics and biology which have not yet been solved and will hardly be solved in the foreseeable future.

Summing up the above brief discussion of the global tendency toward crossbreeding of science and metaphysics, I should like to emphasize that this is not a random phenomenon but a perfectly natural trend showing that the interest in the paradigm of holism in the context of exact sciences has been invariably growing during the past few decades and that research and establishment of generalized mechanisms of synthesis of an indivisible whole from its individual parts is a fundamentally important task.

### 2.4. Hierarchical systems

In his often-quoted article H. H. Pattee [30] wrote, "Evolution requires the genotype-phenotype distinction, a primeval cut that separates energy-degenerate, rate-independent symbols from the rate-dependent dynamics of construction that they control". The use of word 'evolution' makes this statement lose its apparent original meaning. In fact, evolution is directly connected with genetic variability, i.e. with random mutations wherein a gene is a unit of variability. A genotype directly determines and controls a phenotype at the individual organism level, whereas the basis of natural evolution is a feedback that occurs at the population level via natural selection and heredity. Unlike it is the case with ontogenesis, in evolution, genetic symbols are rate-dependent. Essentially, the cause of complexity of life is not so much the diversity of elements, i.e. genes and variants of their phenotype expression, but the diversity of combinations of those elements in populations of organisms. Although molecular biology offers many spectacular successes, it is clear that the detailed inventory of genes, proteins, and metabolites is not sufficient to understand, at the level of additive perception, the cell's complexity [31]. Therefore, the study of those combinations and their hierarchies has become a priority in biology. Structural and functional analysis of hierarchies of genomic and postgenomic data is of primary importance in proteome research [32], studies on metabolism, transcription – i.e. everything that both divides and joins genome and phenome. Thus, Patee's above-quoted statement rather applies to the epistemic cut between potentially measurable properties of a set of system's elements and the unaccountable changeability of the system's properties due to its elements' interaction, given that in reality each unique set of elements correspond to a single specific system with its given unique set of properties. That is why it is so difficult and actually impossible, by using the currently available scientific methods, to calculate that single specific solution that is hidden among an infinite multitude of possible solutions. Ultimately, and in a broader sense, it is all about the epistemic cut between scientific and metaphysical (holistic) views on the nature of things and phenomena and the incomparability of the two views – holistic and reductionist – for the known and the knower. The epistemic cut is of paramount importance, which is especially evident in studies of hierarchical systems.

Any system where there is a parent-child relationship, any ascending or descending series of elements ranked according to their value can be viewed as a hierarchical system. However, upon a thorough consideration it appears that hierarchy, especially concerning biological objects, is an extremely complex notion from the standpoint of philosophical understanding. The fact is that pathways along the branches of hierarchical trees reflect the precise history of node organizations. Individual nodes of a hierarchical tree, even closely adjacent ones, may significantly differ by character and functionality.

A. Koestler [33] referred to the notion of holon which means an entity in a hierarchy that is at once a whole and at the same time a part. Thus a holon at once operates as a quasi-autonomous whole that integrates its parts. Traditional scientific approaches to analysis of complex hierarchical systems are mostly based on reductionism, which suggests that the nature of complex entities can always be understood by breaking them down into simpler or more fundamental components. By doing so it is easy to overlook a certain holon or a number of holons that not only connect but also divide the nodes of a higher than root hierarchical level, which would make them irreversibly lose their hierarchical character. For example, a Chilean economist and philosopher M. Max-Neef [34] has argued that such a complex structure as fundamental human needs is non-hierarchical as human needs are ontologically universal and invariant in nature. However, we all know that any given individual has a specific and distinctly expressed hierarchy of needs. Which one of these views should be the correct one? In fact, both are correct. The problem with Max-Neef's reasoning on a hierarchy of needs is that he disregarded very important holons, psychotypes (according C. G. Yung), which determine the hierarchy of needs of individuals. A hierarchy of human needs is not built from abstract needs, instead it can only exist in connection with an individual whose psychological and physiological specifics and social status determine the pattern of needs. Thus, the conclusion made by M. Max-Neef is valid as an example of utilitarian statistical generalization, while the word 'hierarchy' in this example was used as a trope, a nonliteral reference to inequality, as, for instance, a hierarchy of the three ascending ranks of angels in Medieval Christian theology. Max-Neef's "hierarchy" of needs has nothing to do with the scientific notion of natural hierarchy. There are many other similar examples of use of the word 'hierarchy' which is not science-related, i.e. when the term 'hierarchy is used as an equivalent of assessment of 'more' vs. 'less', 'higher' vs. 'lower', 'more important' vs.'less important', etc., e.g. as in so-called 'dominance hierarchy' which is a form of animal social structure in which a linear or nearly linear ranking exists, or as 'memory hierarchy' in computer science – and does not correspond to the meaning of 'hierarchy' as a scientific term. Oftentimes, a hierarchy is assessed based on analysis of non-commutative properties; or by comparing events that are characterized by conditions and properties that are absent in the preceding node and make other events unavoidable; or by deliberately establishing the nodes of a hierarchical tree and the following re-evaluation of them based on further analysis and consideration; and sometimes, a hierarchical analysis is aimed at simply having a complex system of large parts broken down into a more simple system of smaller parts so it becomes easier for overview and perception. The above-said clearly brings up a question of what is a natural hierarchy and whether it can be scientifically explained as a phenomenon. This problem is the more so important as there is no unanimous opinion on whether or not a natural hierarchy exists at all.

Most research papers that attempt to find out the peculiarities of a natural hierarchy mainly deal with biological objects or phenomena of the biological level of complexity. However, as it will be demonstrated in the experimental section of this paper, a natural hierarchy may exist in any, not necessarily biological, group of objects or phenomena. A. Koestler, in his paper on the theory of self-regulating open hierarchical order (SOHO) [35,36], makes an attempt of highlighting some of the characteristics of the natural hierarchy by describing the properties of holons on the examples of mostly biological objects. Leaving aside the fact that the said description includes 66 positions, which by itself excludes the possibility of providing clearly defined criteria of a natural hierarchy, many of the points of the SOHO theory are equally applicable to both a natural hierarchy and a hierarchy that by no means can be considered as natural. For instance, one (and probably the most significant and interesting of all 66 points) of Koestler's points is: "Every holon has the dual tendency to preserve and assert its individuality as a quasi-autonomous whole and to function as an integral part of an (existing or evolving) larger whole. The polarity between the Self-Assertive (S-A) and Integrative (INT) tendencies is inherent in the concept of hierarchical order, and a universal characteristic of life" [35]. This is certainly correct; however, the same it is true, for instance, for a military hierarchy that is not a natural hierarchy.

It is not difficult to establish that there are two types of hierarchy: descending (D) and ascending (A). In a D-type hierarchy, the root of the hierarchical tree has the most complex organization as compared to higher nodes and end leaves; and conversely, in A-hierarchies, leaves of a hierarchical tree represent more complex structures than the root. For the purpose of analysis, any hierarchy can be considered from the viewpoint of a descending or ascending hierarchy, depending on the analysis vector, i.e. either from root to leaves (d), or from leaves to root (a). An example of A-hierarchy is a phylogenetic tree. The concept of D-hierarchy can be illustrated by the example of an effort to understand how a mechanical clock works by disassembling it into parts. A "working mechanical clock" root produces two nodes: node 1.1, a winding key; and 1.2 , a clock in a working condition but unwound. Node 1.1 is a dead-end node and has no branches. Node 1.2 leads to node 2.1, a clock face that is also a dead-end node, and node 2.2, a working clock which, when wound, can perform the clock's main function, i.e. allow us to approximately determine time by the position of the clock hands. Ultimately, the leaves of Da-hierarchy are: a spring, gears, screws, and other parts that cannot be further broken down. The Da-type hierarchical analysis drastically differs from the Dd-type analysis which can be illustrated by the process of assembly of a clock from a set of individual parts. A Dd-type analysis requires the knowledge of or instructions on performing an assembly. A Da-type hierarchy analysis is based on reductionism, whereas a Dd-type analysis is based on holism. Between the Da- and Dd-approaches, there lies an epistemic cut that may be hard or impossible to overcome: a disassembly is by far easier than an assembly. Therefore, of the four above-said types of hierarchies, the Dd-type hierarchy in any kind of systems is always the most difficult for analysis, especially in comparison with Aa and Ad hierarchies in biological systems. Due to homomorphism of nodes in biological hierarchies, and because a lack of information about some of the intermediary nodes does not significantly affect the hierarchical analysis due to distinct homology between the organisms of successive evolutionary levels, any possible ambiguities in systematics and taxonomy of organisms are mostly local ambiguities that do not interfere with an overall result. In this case, a cognitive process is directed not at the construction of a hierarchy but at the discovery of a natural hierarchy that exists independently from our conscience and has emerged in the process of evolution.

The notions of 'holon' and 'holarchy' [35,36] accepted by the scientific community with much enthusiasm (see, e.g. [37]), essentially, can only serve as an indirect confirmation of the well-known truth that a whole does not equal the sum of its parts. These new terms, having given a rise to speculative tendencies in the study of the 'parts – whole' problem, have not provided any new input into understanding of the universal nature of hierarchy and, particularly, of the mechanism of Dd-hierarchy. Hardly ever can anyone verify, theoretically or practically, A. Koestler's idea that "the concept of the holon is intended to reconcile the atomistic and holistic approaches" [36]. The term 'holon' can be more effectively applied to nodes that can undergo further transformation (evolution). For instance, in the aforementioned example with a mechanical clock, nodes 1.2 and 2.2 can be considered as holons, while nodes 1.1 and 2.1 are not holons. There is at least one more point at where the idea of 'holon' may prove useful. As was already mentioned, in a natural hierarchical system, holons located at different levels of a hierarchical tree should be homological due to their quasi-autonomy. For instance, holons-nodes of the evolutionary phylogenetic tree are homological because each of the objects of that hierarchy represents individual whole cells or organisms made of whole cells. Therefore, the question put forth by H. Pattee [30] "Is it possible for us to distinguish the living from lifeless if we can describe both conceptually by the motion of inorganic corpuscles?" cannot be answered by means of a hierarchical analysis of living forms because the root of the phylogenetic tree of living forms is not a set of macromolecules but a certain hypothetical, most ancient prokaryote-protobiont that had a cellular membrane structure and an established metabolic system. Further, it gave the second branch of the hierarchical tree by having evolved to a eukaryote, a cellular structure in which the DNA replication and ATP synthesis were – with the well-known evolutionary benefit – isolated from the processes occurring on the cytoplasmic membrane, by virtue of respective membrane structures supported by the endoplasmic reticulum network. In any hierarchical system, the degree of heteromorphism among the root and the leaves steadily increases; however, in natural hierarchical systems, the increase of heteromorphism must be strictly successive and regular, and with all the differences between the nodes of various levels, they still must share a certain common feature. In case of the evolutionary phylogenetic tree, such common feature is the cellular structure.

### 2.5. Construction of hierarchies

Construction of hierarchical systems is routinely used in research and technology and represents an important part of human cognitive and intellectual activities. The problem of synthesis of an indivisible whole from a set of parts cannot be tackled without the understanding of mechanisms of construction of hierarchical systems. As for those mechanisms, they cannot be established without the criteria for a natural hierarchy. Certainly, such criteria cannot be determined based on purely theoretical knowledge; however, the vast amount of empirical knowledge accumulated by the humankind appears to be sufficient for establishing the necessary minimum of such criteria and, by that, demonstrating that the modern science does not know the mechanisms of construction of natural hierarchies for the purpose of knowledge acquisition.

Without claiming a comprehensive analysis of all the reasonable criteria of a natural hierarchy, we will point out two most important positions:

**A.** A natural hierarchy is always a closed system representing an indivisible whole. It cannot comprise entities that are not interconnected through branches of the hierarchical tree, i.e. through hierarchical relations. The distinction of a hierarchical system lies in the fact that it can be considered as an element of a higher level system, and that each of its elements, in turn, represents a lower level system. A total interrelatedness is exactly what makes any hierarchical system a closed system. Any hierarchical system can be partially or completely destroyed by the removal of even one of its elements or by the addition of an element that is foreign for a given system.

**B.** Creation of a natural hierarchy is possible only through an objective, uniform and equal impact on all of the system's entities so as to allow the construction of a hierarchical tree to occur based on the principle of self-organization. In a natural hierarchy, the dichotomy of a node is always asymmetrical, unpredictable and depending on the entirety of all the elements and their interrelations. While it is possible to thoroughly study the mechanism of dichotomization, as it has been done, for example, for arterial lines as a system of dichotomous branching [38], one cannot predict the site and time of the occurrence of dichotomy, nor the proportion of the diameter of a new branch to the diameter of the major artery. Dichotomy of nodes is always preceded by a state when the "soon-to-be" groups can completely dissolve in each other and represent a whole. In a natural hierarchy, the formation of branches from an initial node cannot occur through a selective and subjective impact on the system.

Now let us look at the most common techniques used by the modern science for construction of Dd-hierarchical systems, or in other words, for reconstruction of Dd-hierarchy from Da-hierarchy (synthesis vs. analysis). These techniques are referred to as clustering techniques, even though a hierarchy and a clustering are not one and the same phenomenon. Clustering is partitioning of a data set into subsets, i.e. division into groups of similar subjects. Clearly, clustering does not necessarily imply the presence of a hierarchy, whereas a hierarchy always involves clustering. A term

"hierarchical clustering" was coined by S. C. Johnson [39] in 1967. There are two types of hierarchical clustering – agglomerative and divisive. Agglomerative clustering starts with taking each entity as a single cluster and then building bigger and bigger clusters by grouping similar entities together until the entire dataset is encapsulated into one final cluster. Conversely, a divisive hierarchical clustering starts with all objects in one cluster and then subdivides them into smaller units. Divisive clustering methods are not as common as agglomerative. Hierarchical clustering is well covered in science literature, including papers, patents and books (e.g. [40-41]). Here, we will point out those aspects of it which show that the clustering, in its common meaning, performed by the currently available methods, including those which are referred to as "hierarchical clustering", in fact has nothing to do with a natural hierarchy. For that purpose, we will consider the hierarchical clustering from the position of the above-specified criteria of the natural hierarchy: closeness and self-organization.

Hierarchical clustering is usually performed based on pair-wise similarities (or dissimilarities) between the elements of a system under analysis, i.e. by establishing similarity (or dissimilarity) scores for pairs of elements or pairs of groups of elements of the system. A resulting similarity matrix represents an open system, which is expressed in the fact that a removal of any number of elements from an initial system will not affect the (dis)similarities between the remaining elements; likewise, an addition of new elements to a system will not change the relationships between the existing elements. This means that in the process of hierarchical clustering, it is impossible to create a closed hierarchical system, i.e. a system that represents a united indivisible whole. Also, agglomeration and division in the process of hierarchical clustering are determined by algorithms and not by cooperative interactions between the elements of a system, hence they are not a result of self-organization. Moreover, those algorithms allow the use of different techniques for determining the similarities (or dissimilarities) between different elements or groups of elements of the database. Thus performed clustering cannot reflect the natural hierarchy in a system under analysis, as it excludes any possibility for the elements of the system to display their relationships through self-organization.

### 2.6. Bifurcation (dichotomy) as a result of iterative averaging

Previously, we discovered and described a phenomenon [11] that may seem to be contrary to common sense: iterative averaging (both arithmetic or geometric) of properties of the elements of a system leads to dichotomy (bifurcation). By commonsense logic, averaging of the elements of a system should eventually make them indistinguishable. Instead, it causes an asymmetric division of the system into two alternative subgroups. Clearly, this paradox has a general informative value and can provide better understanding of many of the physical processes, natural evolution, and cognitive processes. Since a study of any paradox should start with a solid proof of its existence, which is by far more important than the understanding and explanation of its nature and causes, in this paper we will only focus on the methodology that provides the reproducibility of the said effect and allows its investigation, leaving aside the issues of the metaphysical role of that paradox and its mathematical nature.

We will show on the examples of analysis of real-world and artificial datasets that the discovered effect is universal and true for any kind of system, and that it can be easily demonstrated on any set of data consisting of more than three elements that differ, even very slightly, from each other. The said effect can also be demonstrated on any set of random data points, which distinguishes our methodology for discovery of natural hierarchies in systems under analysis from the so-called hierarchical clustering that cannot be applied to systems of random data points. The proposed method ideally suits the task of construction of natural hierarchies, in an automated mode, by a standardized procedure. First of all, after the very first operation of averaging, any set of data becomes a closed system, which means that none of its elements can be removed from it and no new elements can be added to it, as it would result in distortion of the input dataset. Second of all, the algorithm of iterative averaging provides an absolutely same effect on each of the elements of the system, which is performed autonomously and independently from a human operator. Iterative averaging of the system's elements evokes the processes within the system which are similar to what is considered to be the self-organization processes. These self-organization processes determine the number of clusters in the hierarchical system that emerges from the input dataset. Thus, the said method meets the two earlier defined criteria of natural hierarchy as a certain wholesome structure that is inherent only and only in a given system of data.

After completion of the first cycle of iterative averaging, a dataset under analysis becomes divided into two alternative subgroups of elements, without outliers. The following processing by iterative averaging of each of the successively emerging subgroups ultimately provides a hierarchical system that can be graphically presented as a tree or dendrogram wherein the lengths of branches are proportional to the logarithms of the numbers of iterations that have led to the dichotomy of a given set of elements. Thus, the above-described operational-kind transformations provide images of logical spaces through a visual and holistic representation of the end result of hierarchical construction, which, particularly, allows one to see whether a resulting succession of dichotomies, unpredictable to the human mind, correlates with the human perception of the same system. It is apparent that in such a kind of data processing, even one wrong subdivision into alternative groups, especially at the early stages of the processing, would lead to a completely wrong end result. In the meantime, the processing of data on a hundred of objects can involve dozens of such alternative subdivisions in the course of self-organization of the input dataset. Even the few examples presented in this paper are sufficient for demonstrating that the end results provided by this quite a simple method of data processing are congruent with what is meant by the term 'artificial intelligence'. Certainly, it happens only when an input dataset carries a certain meaning even if it cannot be discovered by any other data processing methods.

The mechanism of transition from the leaves of a hierarchical tree (Dd-hierarchy) to its root, or, in other words, synthesis of an indivisible whole from a set of individual parts is provided by iterative averaging effected by the evolutionary transformation algorithm and does not involve any kind of additional techniques. However, in the course of development of a universal algorithm for the process of iterative averaging of data points, we had to face a number of problems caused by the shortcomings of the currently available techniques for establishing (dis)similarities between objects. First of all, with all the numerous metrics currently available for computation of (dis)similarities (see e.g. [40-41]), there is no logically clear approach to grouping of attributes from the standpoint of applying the most optimal metrics for each type of attributes. For instance, the most commonly and almost universally used metric is Euclidean distances, i.e. distances between objects in a multidimensional space of parameters describing those objects. However, such operations often appear senseless. For example, it is impossible to establish distances between values of such parameters as concentration of a substance or intensity of display of a certain quality of an object. Another example of improper use of Euclidean distance would be a comparison between different levels of household income: in terms of Euclidean distances, the difference between annual incomes of $10,000 and $50,000 is the same as between $510,000 and $550,000, which clearly does not correlate with the actual differences between these values.

In order to normalize and standardize the use of metrics in computation of similarities between objects, we have developed two universal metrics: XR-metric and R-metric [11] (see Methods below) that are applied depending on whether a given parameter reflects a shape or power of objects. Both metrics provide computation of similarities between objects and are normalized from 0 to 1. The XR-metric allows computation of similarities in strict conformity with linear distances between respective objects and provides results that are identical to results obtained for the same objects with the use of Euclidean distances. The examples provided in this paper, as well as our years-long practical application of these metrics demonstrate their rationality and efficacy.

Another problem that has been successfully solved in the course of development of the presented methodology is the effect known as the "curse of dimensionality" [43]. The term "curse of dimensionality" is used to describe a problem that occurs upon the establishing of similarities based on distances measured in high-dimensional space of parameters: the higher is the number of parameters, the less meaning is in similarities computed based on distances between the objects. The algorithm of iterative averaging has removed the curse of dimensionality due to the procedure of hybridization of a set of "monomer" matrices computed separately for each of the parameters [42]. This procedure is described in detail in Methods, section 3.3.

# 3. Methods

Below we describe the algorithm of evolutionary transformation of similarity matrices (ETSM) which is the central point of the methodology of data processing and interpretation presented in this work. All other algorithms provided further in this paper play a secondary, auxiliary role. They contribute to computation quality and accuracy and provide proper visualization of end results.

### 3.1. Evolutionary transformation of similarity matrices

The ETSM algorithm is described by the following equation [11]:

$$[S_{ij}]_{T+1} = Aver((Min([S_{in}]_T,[S_{jn}]_T)/Max([S_{in}]_T,[S_{jn}]_T),n) \quad (1),$$

where "*Aver*" is a geometric (GM) or arithmetic (AM) mean value function, $T$ is the number of similarity matrix transformations according to equation (1), $[S_{ij}]_T$ is pair-wise similarity between objects $i$ and $j$, respectively, after a $T$ number of transformations, and $n$ is the number of objects in a dataset, hence the number of elements in a (dis)similarity matrix under processing. It is important to point up that, as is evident from Eq. (1), the similarity between objects $i$ and $j$, denoted by $[S_{ij}]$, is an indicator that is qualitatively different from a pair-wise similarity between i and j computed in a conventional way based on direct comparison of each object's attributes. $[S_{ij}]$ is computed based on each object's averaged similarity to each of the rest of the system's objects, and therefore it represents a polyvalent similarity between $i$ and $j$ which can be conditionally referred to as 'averaged similarity' (*A*-similarity). Unlike A-similarities, denoted as [*S*], similarities computed in a conventional way based direct assessment of similarities between two objects are denoted as *S* without square brackets. [*S*]-similarity matrices are computed based on initial *S*-similarity matrices. The computation technique is provided in Section 3.3. In computation of $[S]_{ij}$ according to Equation (1), the GM-mode is more practicable, albeit the AM-mode provides results that are qualitatively comparable and non-contradictory to those produced in the GM-mode. Thus, Eq. (1) provides a normalized to the interval 0 – 1 averaging of similarity of each object of the system to all other objects of the system. When using ETSM in the GM-mode for S- matrices based on Euclidean distances which have zero diagonal elements, the first transformation of *S* to [*S*] needs to be performed by using arithmetic means, after which all the subsequent transformations are done with the use of geometric means.

A detailed demonstration of the process of iterative averaging is provided in Section 4.1 on the example of a set of scattered points. Here, in the description of methods, we should like to point out two important properties of the ETSM algorithm.

First of all, the fact of the formation of two alternative groups without outliers does not depend on how the S values were computed. They may be computed based on distances, similarities, dissimilarities, proximities, or any other way of comparison of two objects, including the ways of comparison commonly used by humans or animals for various kinds of assessment of objects. The important thing about the ETSM process is that two objects are compared to each other not directly, as it is done in any other clustering methods, but by their relations to all of other objects in the system under analysis. As soon as after the first transformation of a set of input data, each cell of a square matrix under processing reflects, to a certain degree, the relationships within the whole set of *n* objects, and not only between objects *i* and *j* to whose similarity a given cell corresponds. Thus, as was earlier mentioned, the very first transformation by means of Eq. (1) turns the dataset into a closed system. An addition of a new element to such a system or removal of even one of its existing elements is both senseless and impossible as it would lead to a conflict between the input dataset and its current state system under processing, and the magnitude of such a conflict cannot predicted and pre-assessed.

Secondly, in the course of the iterative averaging, similarities between objects within each of the alternative groups always asymptotically tend to 1 (i.e. to maximum), whereas similarities between objects of different groups asymptotically tend to a certain end value that depends on the values of parameters describing the objects underlying the input dataset. That end value, denoted by Ω is a very important indicator of the process of iterative averaging, which will be demonstrated by us on the examples of various practical applications in the following articles of this series. When the objects within an alternative group are identical, the Ω values may vary from being close to 0 to being close to 1. Upon analysis of datasets describing objects that are very similar to each other, especially when they are described in a multidimensional space of parameters, the 1-Ω values may be diminutively low, up to $10^{-4}$ to $10^{-10}$. Although such low values of Ω do not affect the accuracy and the character of the process of bifurcation, in practice it is more convenient to observe the bifurcation process by using our special 'contrasting technique' which serves as a "magnifying glass".

### 3.2. The function of *Contrast*

The function of contrast, *C*, helps differentiate between Ω and 1, no matter how close to 1 the Ω value may be. The contrast function is designed to attenuate similarity coefficients according to equation (2) [11]:

$$[S]_C = \frac{\exp(\exp S - 1)^{0.082C}) - 1}{\exp(e-1)^{0.082C} - 1} \quad (2),$$

where $[S]_C$ is an *A*-similarity coefficient [*S*] attenuated by the contrast function *C*, and *e* is a natural number. In practice, in the real numbers domain, the contrast function can be applied within the range from 0 to 200, thus allowing any *A*-similarity coefficient in a similarity matrix processed by the method of evolutionary transformation to be represented as either 0 or 1. The effect of the contrast function is illustrated by the plot shown in Fig. 1.

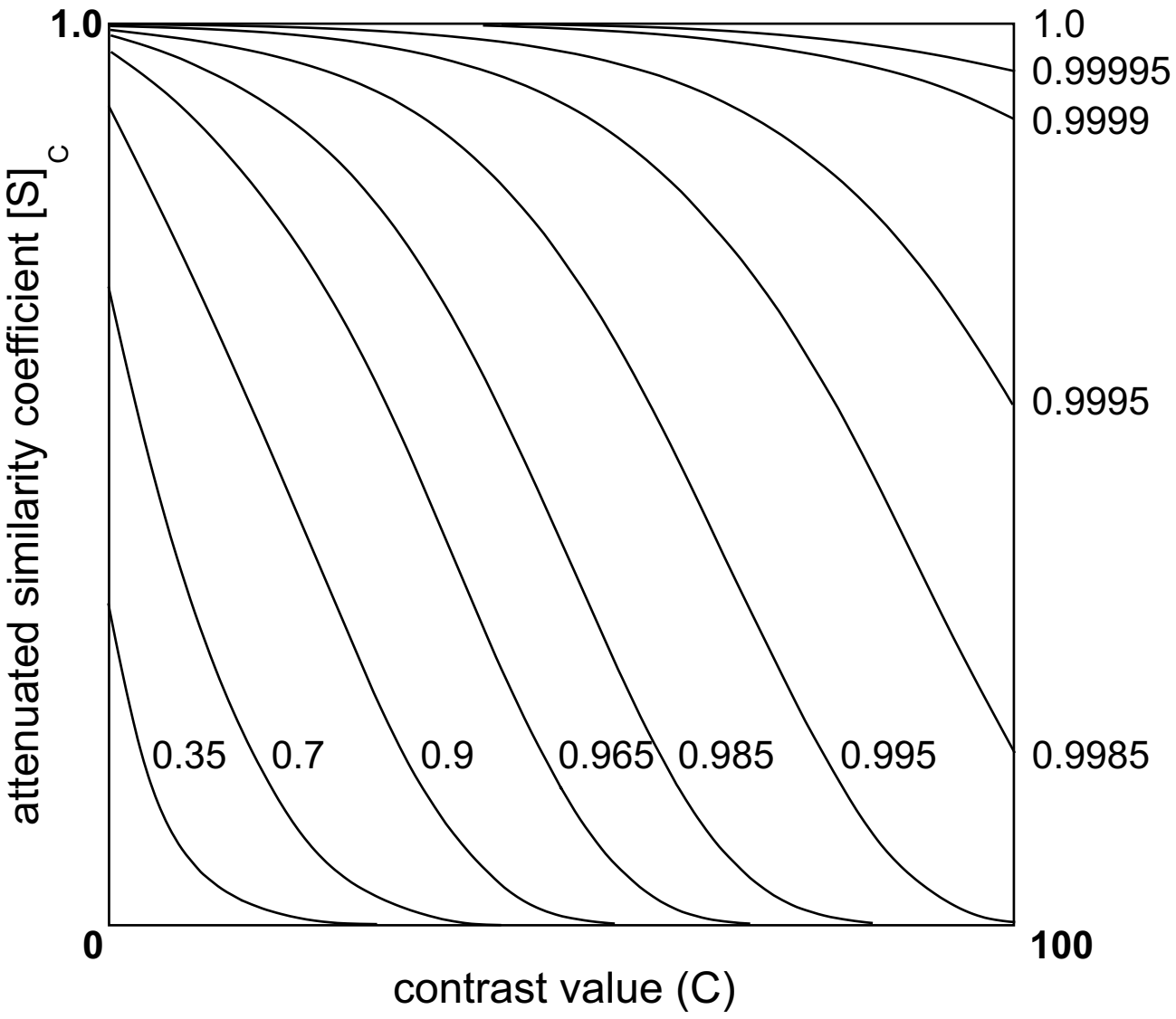

FIG. 1. Dependence of $[S]_C$ from the value of contrast *C* (see Eq. (2). Initial [S] values are shown on each of the 10 curves.

### 3.3. Computation of conventional similarity matrices

As was already mentioned, upon computation of similarity matrices by conventional methods, the increase of the number of parameters describing the objects under comparison, i.e. the increase of dimensionality, leads to a predicament commonly referred to as "curse of dimensionality". The problem is caused by the fact that as the number of parameters increases the distances between objects in the n-dimensional space appear to become progressively lesser parts of the entire volume of the n-dimensional space, thus turning, for instance, Euclidian distances into less and less informative measure of dissimilarities between the objects in a high-dimensional space of parameters. We have developed a method for computation of similarity matrices [42] which eliminates the above-said problem. The method involves computation and hybridization of so-called monomer similarity matrices [42]. Monomer similarity matrices are based on similarities according to one parameter only and are computed for each parameter. Then, for each pair-wise similarity in each monomer matrix, a geometric mean is calculated, which is then used for construction of the similarity matrix for the entire set of objects. Thus obtained hybridized similarity matrix additively reflects objects' similarities based on an unlimited number of parameters. Moreover, it provides a capability to easily change the weight of any of the parameters by changing the share of a respective monomer similarity matrix in the hybrid matrix. It also allows the use of an appropriate metric for each individual parameter.

### 3.4. Metrics for *shape* and *power*

In construction of monomer, hence, hybrid similarity matrices, Euclidian distances are not appropriate as they would transform into city-block metrics. Previously, we have shown that the entire diversity of parameters can be adequately reflected by using only two metrics. One of them is R-metric ("R" for 'ratio') which is calculated by the formula:

$$R_{ij}=min(V_i,V_j)/max(V_i,V_j) \quad (3),$$

where $Vi$ and $Vj$ are values of parameter $V$ for objects $i$ and $j$. Here, similarity values are calculated as the ratio of the lower value to the higher value of the parameter of each of the two objects. Another metric is referred to as XR-metric ("XR" stands for 'exponential ratio') and is calculated by the formula:

$$XR_{ij}=B^{-|Vi\ -\ Vj|} \quad (4),$$

where $V_i$ and $V_j$ are values of parameter $V$ for objects $i$ and $j$, and $B$ (which stands for 'base') is a constant that is higher than 1. XR-metric is designed so that it provides a computation of distances between objects according to desired parameters. Results obtained by using XR-metric fully correspond to those obtained based on Pythagorean Theorem. Unlike Euclidian distances that reflect dissimilarities, R and XR metrics provide similarity coefficients. R-metric is applied to parameters that reflect signal strength, concentration, power, or other intensiveness characteristics. The use of XR-metric is optimal for parameters that reflect a system's shape, a distance between individual points within a system. An important property of XR-metric lies in the fact that variations in the value of constant $B$ do not affect the bifurcation into alternative groups. Thus, by changing the $B$ constant from values close to 1 up to values of the order of magnitude of tens, it is possible to evaluate objects' similarities based on parameters whose values may vary within wide ranges up to many orders of magnitude. Coupled with the method of monomer similarity matrix hybridization, these two metrics provide the advantage of dealing with dimensionless similarity values, which allows for fusion of parameters of any dimensionality, no matter how different and incompatible the parameters may be. Thus, unlike Euclidian distances, the use of XR-metric warrants that neither variations in parameter values nor the increase of the number of parameters or objects can affect the validity of analysis results.

### 3.5. Construction of dendrograms and trees

In construction of dendrograms and hierarchical trees, branch lengths are proportional to a natural logarithm of the number of transformations involved in a complete cycle of asymptotic division of input data into two subgroups. In construction of hierarchical trees, the angles between the branches can be computed according to equation (5):

$$\alpha = \mathrm{ArcCos}\left|-\exp(\Omega - 1)\right| \quad (5),$$

where $\alpha$ is the angle between the hierarchical tree branches that represent each of the two subgroups, a $\Omega$ is the limit value of similarity between two subgroups of objects which is reached at full completion of the formation of two alternative subgroups. The $\alpha$ value varies from 0 degrees when similarities between two subgroups equal 1, up to 180 degrees when similarities between two groups equal 0.

### 3.5. Software

All the analyses reported in this paper were done with the use of computer program *MeaningFinder 2.2* (Equicom, Inc).

# 4. Experiments

In this section, we provide a number of examples of analysis of datasets to demonstrate the mechanisms of application of our methodology to synthesis of an indivisible whole from its parts. In other words, we will demonstrate how it provides, in an autonomous computing mode, a thorough investigation into complex datasets on objects presented in a multidimensional space of attributes. The examples include both complex ones, such as a comparative study of the climates of 100 cities of 42 U.S. states based on 108 meteorological parameters, and relatively simple 3-D spaces of scattered points. As was already mentioned, the uniqueness of the presented methodology is manifested in its capability to handle an input dataset as a closed self-organizing system wherein all of its elements interact with each other according to a certain intrinsic logic that does not depend on the will of a data analyst or programmer. This phenomenal feature of the ETSM methodology is evident from the few examples presented in this paper.

When a programmer or a team of programmers develop a data processing flow, it is always a set of additive operations, no matter how many steps it may include and whatever complex mathematics may be involved in each of the individual steps. In a resulting program, these steps may be performed consecutively or concurrently, but they can always be sorted out, i.e. separated from each other. Certainly, they are connected by a certain logic that underlies a given program. Except for intuitive reasoning where information is usually supplied in the form of entangled blocks of loosely connected fragments, a similar approach is used in classical science procedures utilized in the course of solving certain research tasks. As far as the method of iterative averaging is concerned, it does not have any counterparts in the techniques used by the human mind in the process of cognition: here, in order to understand a system or an event or a phenomenon, we take all of its individual elements, bond them together into an indivisible whole and then subject it to processing, which represents an absolutely and ultimately holistic approach. This is the first most important peculiarity of the ETSM method. The second important peculiarity of the method is that the ETSM data processing represents a series of absolutely identical operations performed according to one and the same Equation (1). And finally, the third peculiarity of the method lies in the fact that the end results of analysis by the ETSM method are neither reducible to, nor deducible from the original input data.

The ETSM method represents a new, heretofore unavailable method of knowledge discovery through a sort of "matrix reasoning". It can be used to solve any information analysis tasks (as was earlier mentioned, hybridization of monomer matrices [42] removes any limitations in terms of the number of parameters describing the objects under analysis). Datasets that lack any logical meaning or the presence of several overlapping centers of conflicting information in an input database will present a problem, as well as they do for traditional methods of cognition. In the following publications of this series, we will demonstrate some of the techniques that allow the ETSM method to correct such problems.

**4.1. Analysis of scattered points**

To demonstrate a result of the iterative action of the algorithm ETSM described by Eq. (1), we will refer to an example illustrated on Figs. 2a – 2b. Fig. 2a shows a set of 36 scattered points which clearly look like four distinct groups of points; we have labeled the four groups as A, B, C, and D, and marked out two points in each group: **a** and **a1**, **b** and **b1**, **c** and **c1**, **d** and **d1**. Further, pair-wise *A*-similarities between the points are denoted by: **aa** for **a** and $\mathbf{a}_1$, **bb** for **b** and $\mathbf{b}_1$, and so on; **ab** for **a** and **b**, and so on.

The input S-similarity matrix of 36 scattered points was computed based on Euclidean distances and was further processed according to Eq. (1) using arithmetic means for the first transformation of $S$ to $[S]$ and geometric means for subsequent transformations (see Methods, section 3.1). After the first 300 transformations, similarity coefficients for pairs **aa**, **bb**, **cc**, **dd**, **bc**, and **cd** appear to equal 1 with an accuracy of up to the 7th decimal place, whereas similarity coefficients for pairs *ab*, *ac* and *ad* (i.e. Ω) equal 0.89459. Upon completion of 500 transformations, changes in the Ω value occur only in the ninth decimal place. By applying the contrast function (see Methods subsection III.2.) at C=80, the result of the evolutionary transformation of the similarity matrix of this data point set can be presented, practically, by 0 and 1, thus providing a complete separation of 16 points of group A from the rest 20 points of groups B, C and D. The evolutionary transformation of 20 points of groups B, C and D after their separation from the points of group A results in formation of two loci: 1) 8 points of group B, and 2) 12 points of groups C and D . Upon the evolutionary transformation of the 12 points of the second locus, 8 points of group B get separated from 4 points of group C и 4 points of group D. Fig. 2b shows an iso-hierarchical picture of the process of division of the set of 36 scattered points: the higher was the number of the transformation-division-transformation cycles required for formation of a certain node of the hierarchical tree, the darker is the area of points joined in that subcluster, and vice versa. Iso-hierarchies, as well as hierarchical trees and dendrograms provide visualization of hierarchical structures constructed through the use of the ETSM algorithm.

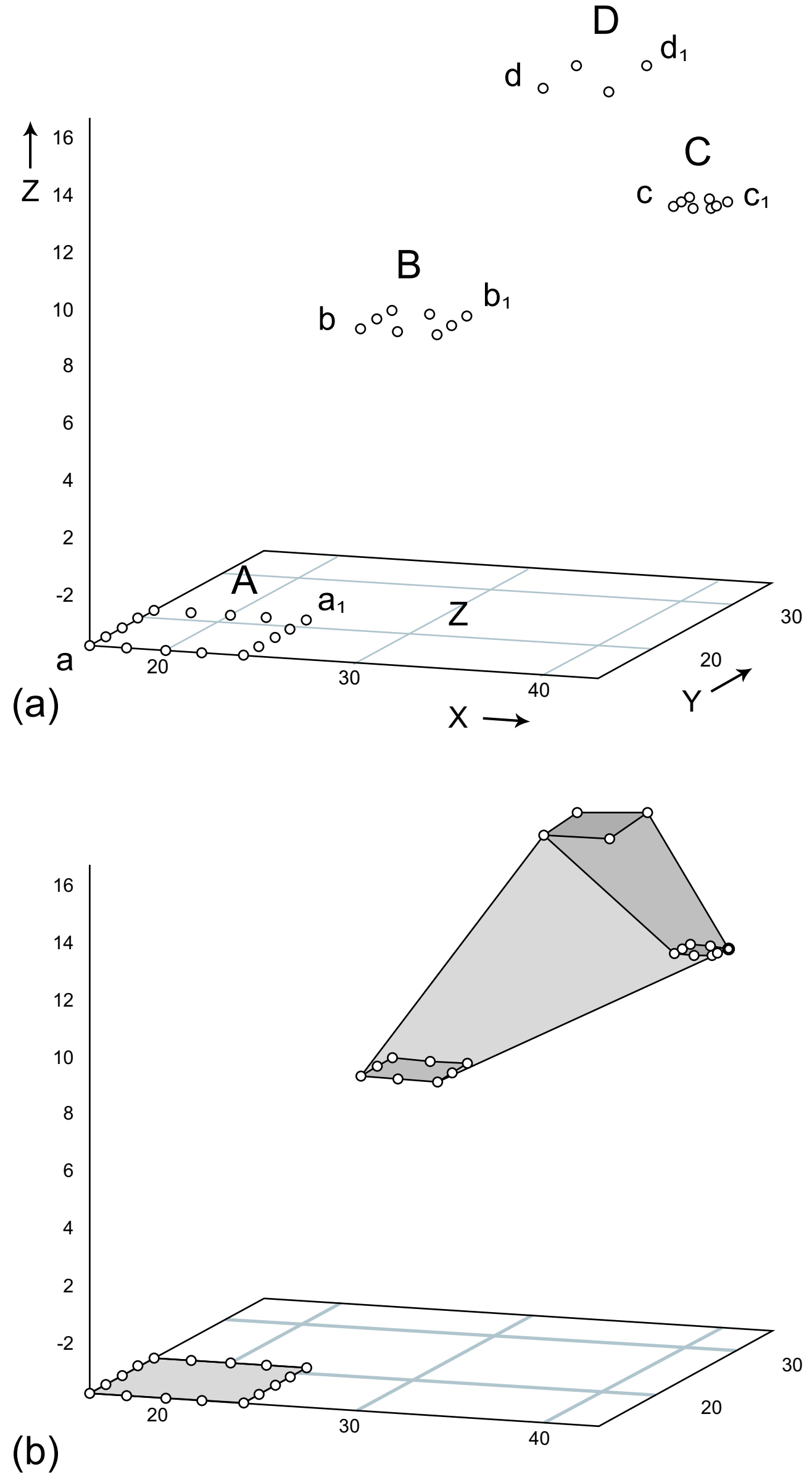

FIG. 2: 3D space of 36 scattered points before and after ETSM processing. a) 36 scattered points located as four groups: A, B, C and D. b) Same 36 scattered points after ETSM processing presented in the form of iso-hierarchies: the greater is the affinity between the objects, the darker is the shading of the plane that connects the objects.

Figs. 3a – 3c demonstrate the nonlinear dynamics of the evolutionary transformation, at the contrast value of 50, resulting in subdivision of the *A*-similarity values. The result shown in Fig. 3a was obtained by applying Eq. (1) in the AM-mode, whereas the results presented in Figs. 3b and 3c were produced in the GM-mode (see Methods subsection III.1.). As is seen upon comparison of Figs. 3a and 3b, the dynamics of the evolutionary transformation are qualitatively same in both of the modes, with only slight quantitative differences. A comparison of Figs. 3в and 3c shows that the dynamics of the separation of the group B points from groups C and D has the same regularities that were observed in Fig. 3b upon separation of the group A points from the rest of the points.

The above-demonstrated method for fully unsupervised hierarchical analysis significantly differs from the heretofore known clustering methods including those that are commonly referred to as "unsupervised". Firstly, upon the very first transformation of the similarity matrix in the above example, the objects of the system under analysis become objects of a close (i.e. isolated) cooperative system that immediately starts evolving into two separate closed subsystems representing the first two branches of the system's hierarchical tree. Secondly, due to the above-said peculiarities of the ETSM algorithm, no object can have affinity to both sub-systems, and therefore, after a certain number of transformations, no outliers, i.e. objects that tend to both of the two loci, are left. Thirdly, the number of nodes of the hierarchical tree (see, e.g., Fig. 2) corresponds to the number of successive transformation-division-transformation cycles, each of which results in formation of two loci, whereupon each newly formed locus is subjected to ETSM; thus, the number of nodes depends solely on the innate structure of an input data system and by no means is set at the analyst's discretion, as is the case with, for instance, k-mean clustering. Finally, unlike all of the commonly accepted clustering methods that are aimed at organizing the diversity of a system's objects by sorting them, this method provides the evolution of the diversity of a system's objects as a whole, leading a system to transformation into two opposite subsystems. The above-described ETSM process represents a peculiar combination of convergent and divergent evolution of a complex system's objects, which is stimulated by the averaging of the objects properties.

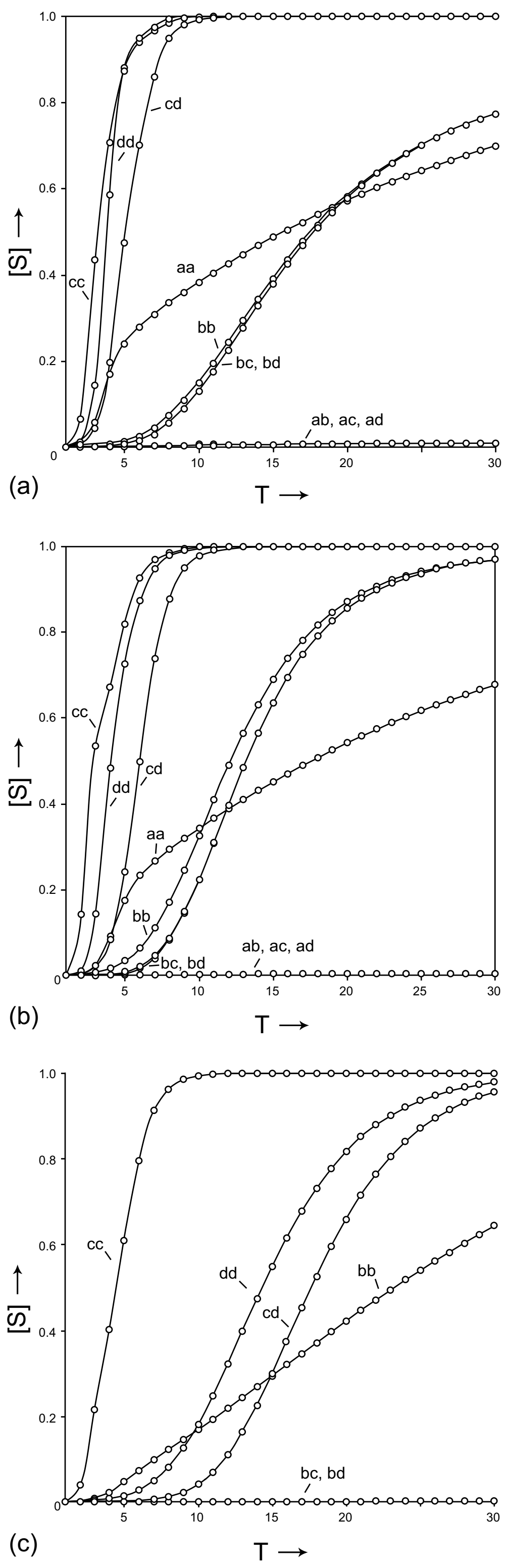

FIG. 3: Dynamics of evolutionary transformation of similarity coefficients of 36 points (shown in FIG. 3a) in the course of iterative processing by the ETSM algorithm. A-similarities [S] are shown to change depending on the number of transformations (T). Inter-group similarities were determined based on points "a", "b", "c" and "d", and intra-group similarities were determined based on pairs "a" and "a1", "b" and "b1", "c" and "c1", "d" and "d1" and are denoted as "aa", "bb", "cc" and "dd", respectively. Dissimilarity matrices were computed based on Euclidian distances. The value of contrast was 50 (see Methods subsection III.2.). 3a) Evolutionary transformation of 36 scattered points in the AM-mode; 3b) Same in the GM-mode; 3c) Evolutionary transformation of 20 points of groups B, C and D in the GM-mode after separation from the group A points.

It is important to point up the following. Evolutionary transformation of any similarity matrix, regardless of the number of objects and parameters, occurs according to one and the same scenario. The above-demonstrated examples of nonlinear dynamics of ETSM (Figs. 3a –3c) clearly indicate that there may be multitudes of similarity matrices whose evolutionary transformation will produce one and the same result. Thus, unlike chaotic nonlinear dynamics, the ETSM nonlinear dynamics results in not the increase of diversity but, on the contrary, unification of the objects of a complex system and in dichotomy of trajectories of variations in multiplicity of similarity matrices.

The above-provided simple example of analysis of a dataset of scattered points demonstrates a mechanism of unsupervised construction of a hierarchical system that, in our opinion, meets the criteria for a natural hierarchical system. This mechanism can be used for discovery of hierarchies in any kind of database, including 3D and multidimensional systems, as well as spatial-temporal systems. The XR-metric (see Methods, III.4) coupled with construction of similarity matrices through hybridization of monomer matrices (see Methods, III.3) removes the problems that occur in such computations upon the use of Euclidean distances. As is seen from Fig. 4a, upon the increase of the number of scattered points to 115 and the use of Euclidean distances, the division into hierarchical groups is partially incomplete. The use of Euclidean distances in multidimensional systems provides even worse results. As is seen from Fig. 4b, the XR-metric is free from that drawback.

## 4.2. Hierarchical analysis of randomized datasets

It is important to realize that hierarchical grouping (subdivision) of mathematical points even in a 3D space depends on completely unpredictable factors. Those factors are determined by the relationships between all of the elements of a system under analysis, i.e. by their overall cooperative interactions. Certainly, those interactions are inherent, although in a hidden form, in any dataset under analysis, but they become detectable only after the first cycle of iterative averaging which transforms the initial dataset into a closed system, i.e. a system in which the principle of holism is manifested in its full.

It should be emphasized that the ETSM algorithm is not a modeling tool or a tool that enables a data analyst to choose the most optimal solution among a number of possible solutions. This is the fundamental distinction of the ETSM algorithm from all the algorithms for data clustering. Data analysis performed through the iterative averaging procedure provides only one and final result. That result will be logically meaningful only in case when input data inherently contain a certain logical foundation, a certain meaning that needs to be extracted and presented for understanding. The above factor is essential for the understanding of the unique capabilities of the 'matrix reasoning' technology as a new approach to data processing which seems to have come up very close to that vague concept that is generally referred to as 'artificial intelligence'.

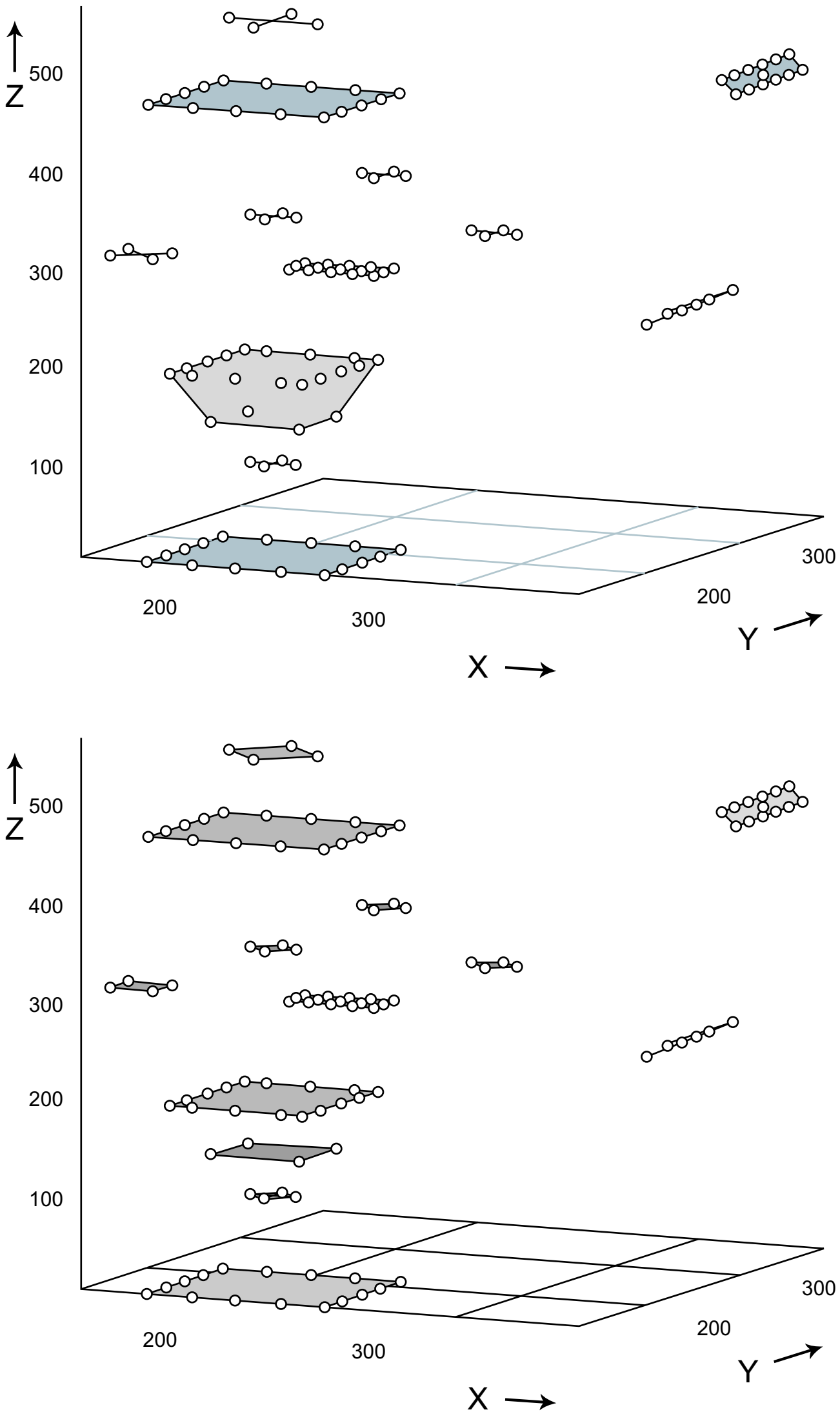

FIG. 4. Iso-hierarchies of 115 scattered points obtained with the use of the ETSM algorithm: 4a) based on dissimilarity matrix computed by using Euclidean distances between the points in a 3D space and 4b) based on hybrid similarity matrix computed with the use of XR-metric (B = 1.1).

Clearly, it is impossible to provide a theoretical proof of the unique capabilities of 'matrix reasoning' since a conceptual plan contained in a given dataset cannot be *apriori* calculated, nor can they be proven by merely demonstrating a limited number of examples of its practical application as we do it in this paper. The capabilities of this method can be evaluated only by trying it in processing of various kinds of information, using the techniques described in this paper. In this section, we will provide an example *ex contrario* by analyzing a dataset which, by definition, cannot carry any meaningful information. The example below is based on a set of randomized data.

Fig. 5 shows three hierarchical trees obtained by ETSM-processing of data tables for 500 objects described by 500 parameters whose values in the range of 1 – 500 were generated by a random number generator. A dichotomous subdivision will certainly occur as a result of iterative averaging even if among the 500 objects there is only one pair of distinguishable objects.

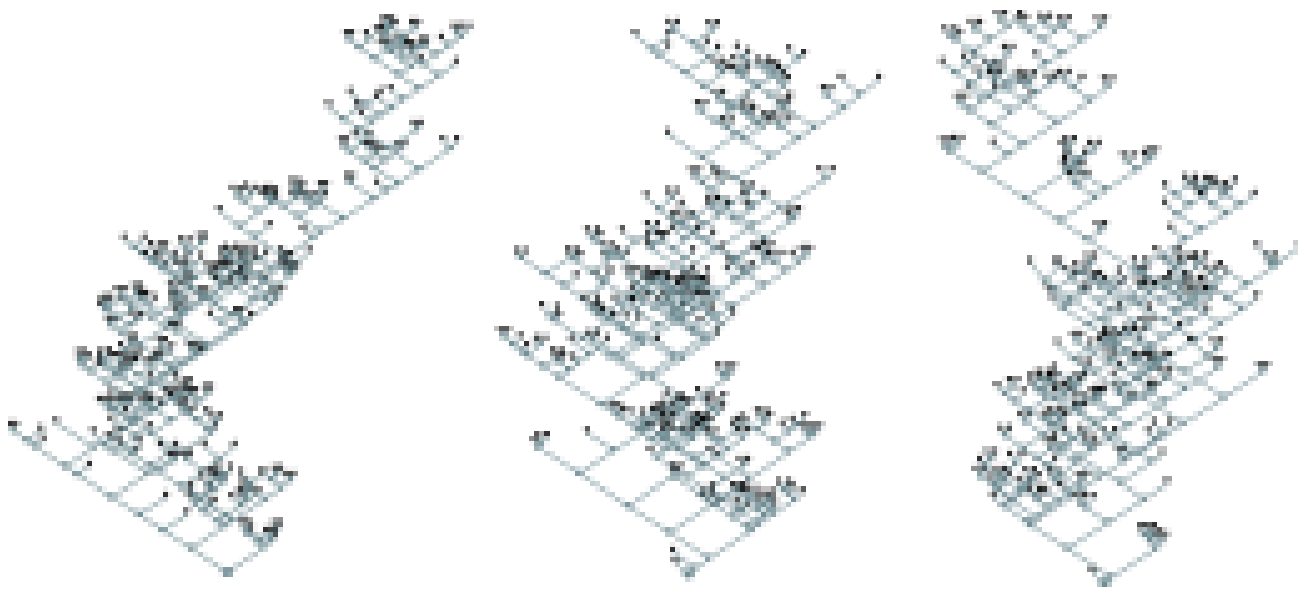

FIG. 5. Hierarchical trees obtained through iterative averaging of three datasets, each of 500 objects described by 500 parameters whose values in the range from 1 to 500 were produced by a random number generator. Similarity matrices were computed by using the XR-metric (see section III, Methods).

Therefore, there is nothing unusual in the fact that a set of randomized data can actually form hierarchical trees. However, as one can assume, hierarchical trees of randomized sets of the same number of data points will not greatly differ in the number of nodes and leaves and cannot convey any meaningful information, and, most importantly, it is practically impossible to obtain two identical hierarchical trees as any randomized dataset will produce a unique hierarchical picture and can be easily identified. These assumptions proved to be true when we analyzed about 200 randomized datasets. Each of the resulting hierarchical trees was different. The three hierarchical shown in Fig. 5 have different shapes and consist of 39, 29 and 31 nodes and 304, 310 and 309 leaves, respectively.

### 4.3. Comparative analysis of climates of U.S. cities

It follows from the above provided description of the ETSM methodology that formation of natural hierarchical structures for any given dataset processed by the iterative averaging algorithm always occurs according to one and the same scenario involving the same standard procedures since our system of data processing is engineered to take scalability into account. This is achieved through the way we construct similarity matrices - by hybridization of monomer similarity matrices computed for each individual parameter. Thus, the resulting hybridized similarity matrices are based on dimensionless similarity criteria. The analysis of thus processed data is essentially based on comparison of the positions of objects in the rows, normalized within a range of 0 to 1, according to individual dimensionless characters. This allows a concurrent processing of an unlimited number of parameters expressed in different units and thus eliminates the necessity of reducing the number of parameters by selecting the most representative ones, for which mathematical statistics usually employs a discriminant analysis [44].

In this section, we will demonstrate quite a complex example of multi-parameter computations by the ETSM method. We processed comparative climatic data provided by the U.S. National Climatic Data Center [45] for 100 U.S. cities of 42 states, including the following 108 climatic characteristics based on thirty-year averages for each parameter: morning and afternoon values of relative humidity, in per cent, for each month of the year (the total of 24 parameters); relative cloudiness, in per cent, based on average percentage of clear, partly cloudy and cloudy days per month (the total of 36 parameters); normal daily mean, minimum, and maximum temperatures in degrees of Fahrenheit (the total of 36 parameters); and normal monthly precipitation, in inches (the total of 12 parameters). All data are based on multi-year records from the year 1970 through 2000. A similarity matrix was constructed according to the method for hybridization of monomer similarity matrices with the use of XR-metric at B=1.5 (see Methods sections 3.3. and 3.4.).

The said computations involve a total of 10,800 data points. Taking into account that each data point represents an average based on 30 measurements, a total of the underlying data points is over one-third of a million data points. It should be also taken into account that deviations from average values significantly vary from parameter to parameter and from location to location, thus making the input dataset an extremely chaotic system. A discovery of a reasonably distinct correlation between the parameter values and geographic locations would be impossible by applying any of the currently known data processing methods. The difficulty of establishing correlations in such a system is caused by the fact that the dynamics of variations in values of meteorological parameters is extremely nonlinear, and each parameter's mean values for any particular year are greatly influenced by that year's meteorological specifics.

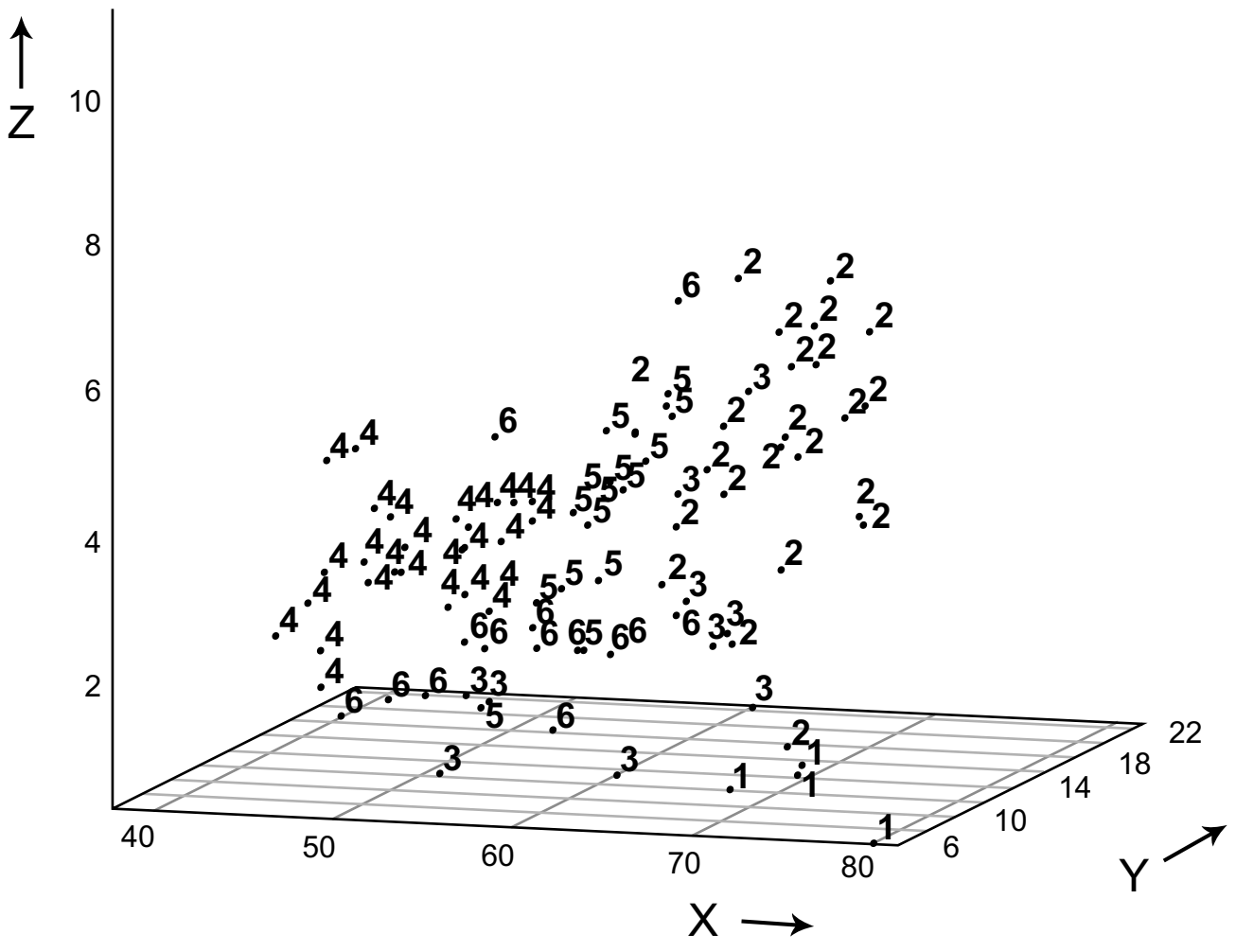

FIG. 6. Relations between 30-year (1970 – 2000) averages of monthly (March) normal means of daily maximum temperatures in °F (*x*), cloudy days per month (*y*) and precipitation (in inches) (*z*) for 100 cities of 42 states of the USA.

Even a simplified example based on only three meteorological parameters for only one month, March, (see Fig. 6) shows that there are no apparent correlations between geographical locations and the multi-year averages of the three meteorological parameters: normal daily maximum temperature (°F), cloudy days per month, and normal monthly precipitation (in inches).

Fig. 7 shows a dendrogram of 100 U.S. cities which was obtained based on all 108 above-said parameters processed by the ETSM-algorithm in the unsupervised mode. As is seen from the dendrogram, cities located in same states appear to be in the same subclusters, and there are six distinct climatic groups of states.

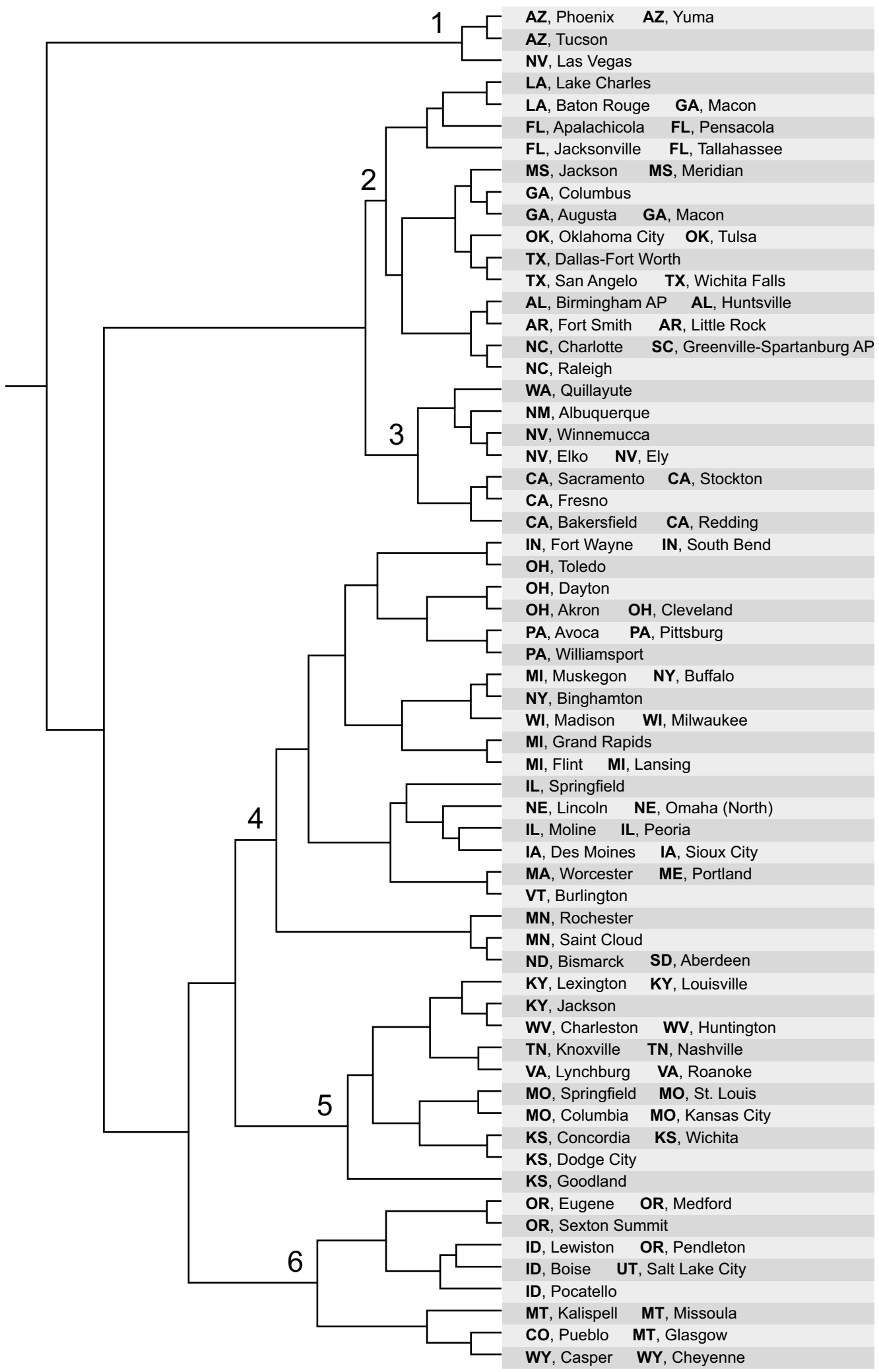

FIG. 7. Dendrogram of 100 cities of 42 states of the USA, produced by the method of iterative averaging, based on 108 climatic parameters

(specified in the text). The resulting subclusters are indicated by numbers 1 through 6.

On the USA map in Fig. 8, we have labeled each of the states involved in the analysis by numbers 1 through 6 in accordance with the grouping shown by the dendrogram in Fig. 7: as is seen, the six groups cover the totality of groups of northern, central, southern, and western states, thus demonstrating perfect dechaotization of the input data. The fact that the obtained grouping is not a result of mechanical mathematical operations is especially obvious from viewing group 5 that joins together the central states as a narrow layer between the northern (group 4) and southern (group 2) states.

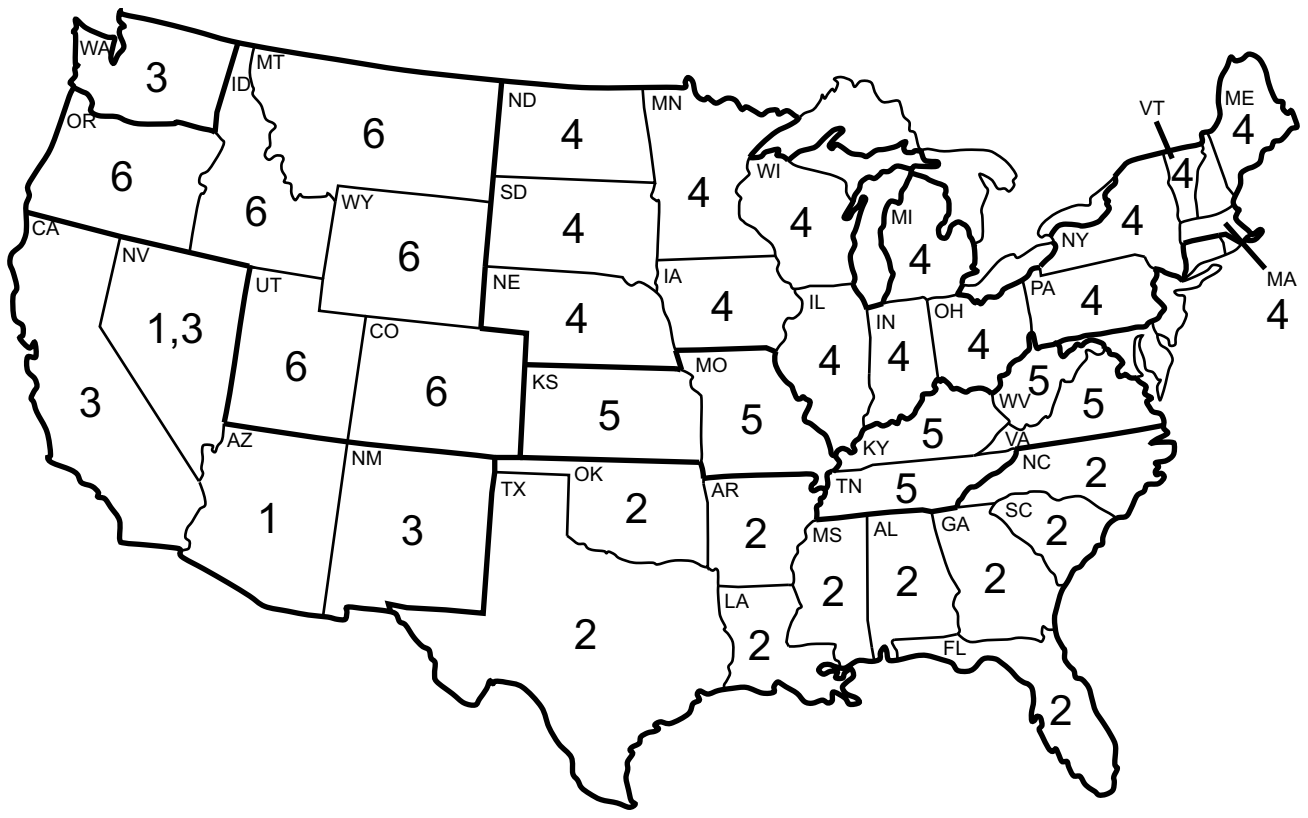

FIG. 8. Grouping of 42 states based on 108 climatic parameters. The states are labeled by the numbers that correspond to the group numbers in the dendrogram of the cities located in the respective states (Fig. 7).

Thus, we have demonstrated, on an example of a real-world dataset, that it is possible to synthesize an indivisible whole from a set of parts. In this example, the indivisible whole is the dendrogram of the cities according to their climatic peculiarities, and the U.S. map showing the U.S. states grouping based on the said character. Taking into account the high number and diversity of the involved parameters, this grouping is definitely not a result of any accidental coincidences.

### 4.4 Hierarchical analysis of population pyramids

Population pyramids are a graphic way to show the age/gender composition of a population and its age/gender structure (usually composed of 5-year cohorts) [46]. They look like nearly symmetrical bell curves and represent a basic tool in demography. The shape of a population pyramid is basically a result of birth, death, and migration rates. However, quantitative characteristics of evolutionary population pyramids significantly depend on various factors: ethnic, socioeconomic, ecological, climatic, political, and others, as well as on many events that are extremely difficult to evaluate and take into account. Therefore, population growth results always contain a great deal of unpredictable, chaotic components and dynamic instabilities. Even though population pyramids are believed to be a self-explanatory reflection of the state of a country's population, the use of population pyramids in correlative analysis is highly complicated. An example of demographic analysis presented in this section provides one more compelling demonstration of the efficiency of iterative averaging in extraction of information from highly complex databases.

We processed demographic data on 72 countries, including 50 demographic parameters according to U.S. Census Bureau data for the year 2000 [47]. The 50 parameters are: population pyramid sections reflecting percentages of various age groups (in 5-year cohorts) from 0 to over 80 year-olds in the total male and female populations, respectively, (total of 34 parameters); birth and death rates per 1000; life expectancy at birth; infant deaths; total fertility factor (total of 5 parameters); and dynamics of population growth in various years (in 1980, 1990 – 1999) compared to the year 2000 (total of 11 parameters). The list of 72 countries includes: 34 countries with predominantly Muslim populations; 21 European countries of the former Soviet bloc and former USSR republics with predominantly Christian populations; Israel, with predominantly Jewish population; and 16 European countries with free market economy and predominantly Christian populations. A similarity matrix for 72 countries was computed by using R-metric (see Methods, subsection III.4). As well as in all of the above-provided examples, the entirety of the data was subjected to automated unsupervised processing by the ETSM algorithm

The fact that the database under analysis was highly heterogeneous and extremely difficult for extraction of information is demonstrated by a small example on Fig. 9. It shows a dependency plot of two parameters that are very similar in their nature: the portions of males in the age of 30-34 and 35-39 in population pyramids of 72 countries. As is seen in Fig. 9, even such closely related parameters as male portions in the age groups of 30-34 and 35-39 of 72 countries do not display any definitive correlations.

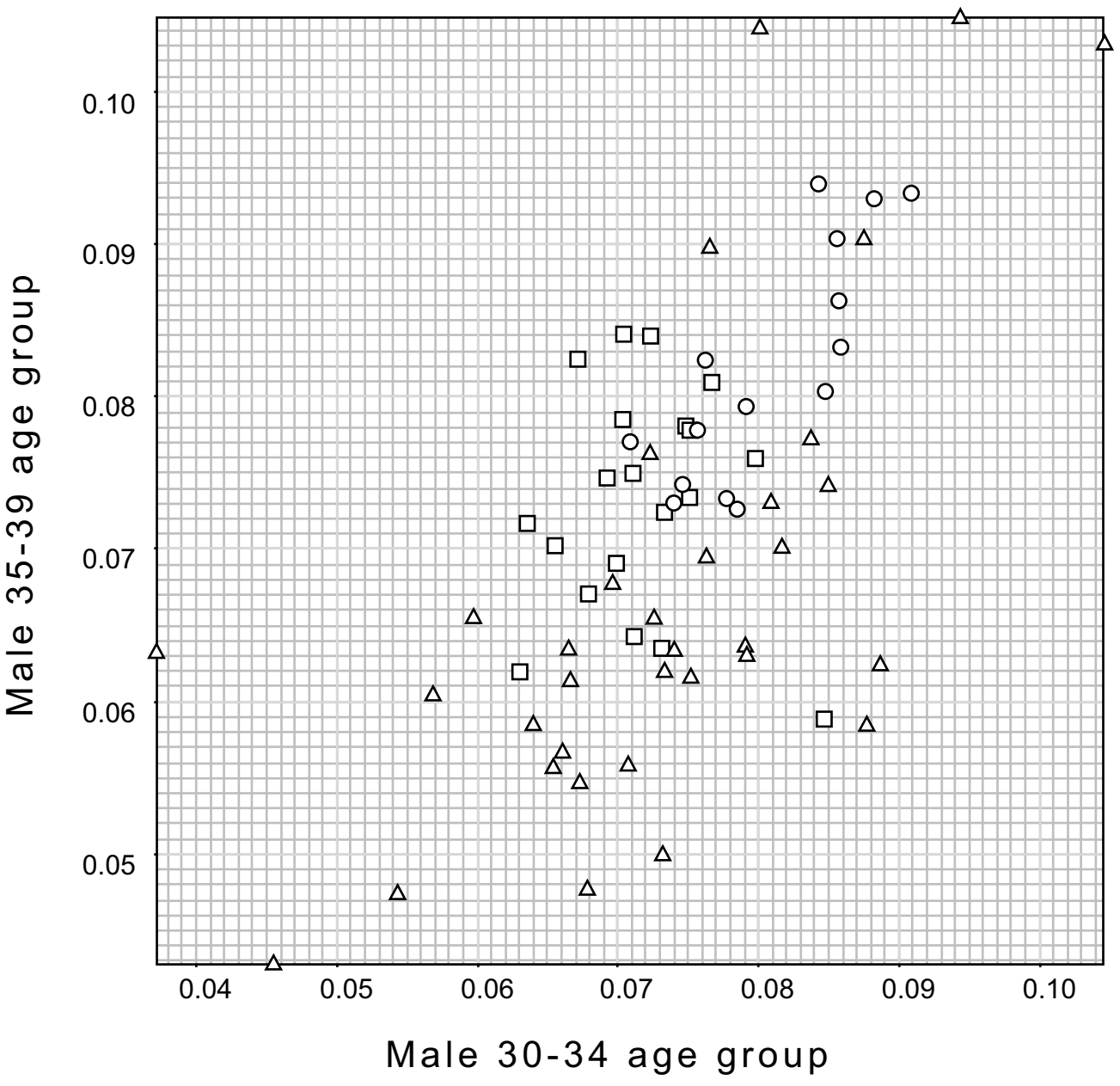

FIG. 9. Relations between male portions in the age groups of 30-34 and 35-39 of populations of 72 countries. Triangle-shape data points correspond to Muslim countries; squares, to the countries of the former Soviet bloc and former USSR republics; and circles, to the countries with historically free market economy.

An ETSM-produced dendrogram presented in Fig. 10 shows three distinct loci that exactly correspond to the above-indicated three groups of countries. It clearly distinguishes Muslim countries from the rest of the countries. The latter, in their, clearly show two groups consisting of: former Soviet bloc countries, and the countries with historically free market economy. Israel appears to be in the same subcluster as the former Soviet bloc countries, which seems to be logical, given a high percentage of immigrants from those countries in the total population of Israel, as well as certain peculiarities of its social policy. There are some other interesting "coincidences" in the grouping of the countries: e.g. a group including all the seaside countries of southern Europe (Italy, Portugal, Spain, and Greece); a group of countries that are different demographically and economically but are close geographically, historically and culturally (Jordan, Yemen, Iraq, Syria, Djibouti, Oman, Saudi Arabia); countries of the Maghrib; countries with oil-based economy; Muslim countries of South-East Asia, including former USSR republics in Middle Asia; a well-defined grouping of Russia, Ukraine, Belarus and the Baltic states, etc. This data processing example is not aimed at interpretation of the obtained results, its purpose is to demonstrate that the above-presented unsupervised ETSM-analysis of demographic data has successfully identified three distinctive loci in the group of 72 countries and discovered certain natural logic (in the form of religious, cultural, political, economic, geographical, and other correlations) in the grouping of countries within the three main loci.

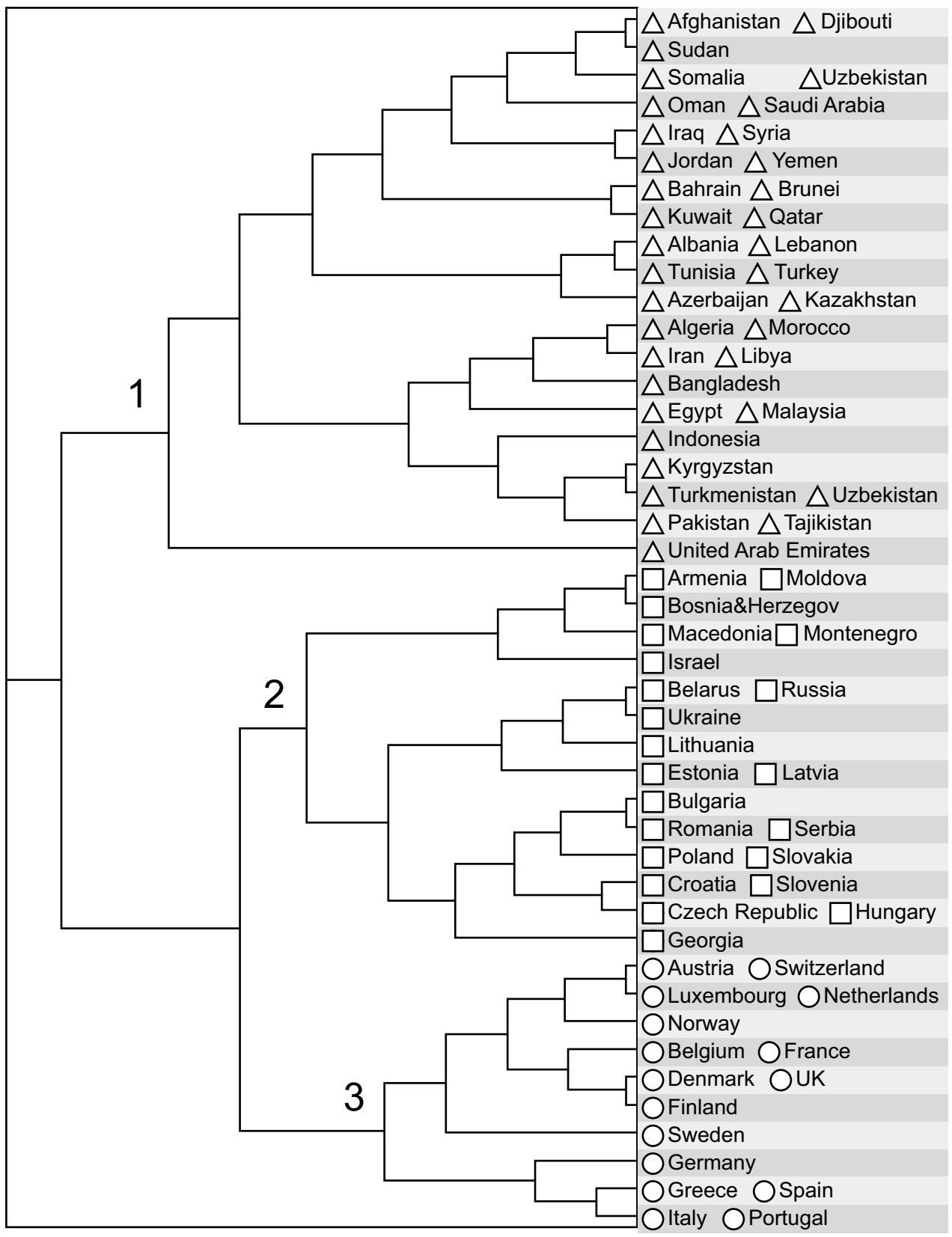

FIG. 10. Dendrogram of 72 countries, obtained by BfC-clustering based on 50 demographic characteristics.

All of these details are the evidence of the input data successful dechaotization as a result of processing them with the ETSM algorithm. A simple mathematical process applied to an entirely entangled data array has produced a crystal clear picture of similarities and differences between the countries described by the data under analysis, showing that the quantitatively measured parameters perfectly corroborate common knowledge of the kind that cannot be reduced to plainly quantitative characteristics.

## 5. Some concluding remarks

The foregoing is a description of a fundamentally new approach to data processing, which has no analogs in the nowadays information science. The algorithmic basis of the presented method is fairly simple, and a reasonably informed reader who possesses basic programming skills can easily, even without the use of a computerized implementation of the method, such as, e.g. MeaningFinder 2.2, that iterative averaging concurrently applied to all of a system's elements always results in subdivision into two alternative groups. However, unlike the final result, the process of its achieving is very difficult to track down and monitor. Dichotomization, being a highly non-linear process, is practically impossible for visualization, which certainly can slow down the assimilation of this very promising technology. A specialist educated on linear cause-effect principles will definitely have trouble perceiving the fact that in order to have a system subdivide into certain meaningful substructures, all of its elements need to be mixed into an indivisible whole

Another quite puzzling peculiarity of the iterative averaging algorithm is the fact that, as is seen from the earlier provided practical application examples, despite its relative simplicity, this algorithm displays an expressly intelligent response, which would not be totally unexpected should it be some kind of a highly sophisticated computer program and not merely an algorithm based on a mechanical repetition of one and the same operation.

And finally, there is a third question that cannot be avoided. We have demonstrated that the averaging of a system's elements results in formation of two heterogeneous groups, instead of homogenization of the system as one would expect based on common sense. This equals an assertion that there should be a certain underlying physical principle that, obviously, should be discoverable through adequate physical methods, for instance, in the course of studies on turbulence, quantum-mechanical effects, biological evolution processes, etc.

Each of the afore-mentioned issues certainly require in-depth consideration and explanation; however, the purpose of this paper was to provide a detailed description of the new technology and to emphasize that the principle of holism, on which this technology is based every whit, is not just a tool for philosophical understanding of reality; instead, it is a reality of modern science and technology.

Even if in addition to a few examples provided in this paper we would have given a few dozens of examples from our decades-long work with the algorithm of iterative averaging, it would not add an iota of further knowledge that has yet to be discovered on this phenomenon, as the discovery of such knowledge would require a different kind of investigation. The fact that the algorithm of iterative averaging provided in this paper represents the most natural way of transition from linearity to non-linearity in the real world is obvious. This fact has yet to be investigated from the standpoint of mathematics.

The remarkable intelligence potentials of the algorithm of iterative averaging, only in part demonstrated in this paper, also require special studies. Apparently, the key factor here is a natural hierarchy whose manifestation is facilitated by the algorithm. Although a strictly scientific and precise definition of natural hierarchy is probably impossible to provide, we pointed out two most significant characteristics of a natural hierarchy: a system's closeness and capability for self-organization. If one would apply the algorithm of iterative averaging to a database of words with known numbers of locations characters, it would act only as a search engine sorting the words by length and character composition, since there is no natural hierarchy in such a database: the meanings of words are not a natural result of their character composition; instead, they were conventionally assigned to words by users of a respective language.

On a final note, it should be added that the method of data processing by iterative averaging offers a possibility of various implementations in knowledge discovery, the most interesting of which we will describe in the following papers of this series.

**Acknowledgments**

Thanks are due to Oleg Rogachev for his work on the software implementation of the presented methodology. I am also grateful to Michael Andreev for the graphic visualization work.